\documentclass[aps,prd,twocolumn,eqsecnum,amssymb,amsmath,showpacs,a4paper, superscriptaddress]{revtex4-2}

\usepackage{bm}
\usepackage{hyperref}
\usepackage{framed}
\usepackage{subfigure}
\usepackage{hyperref}   
\usepackage{graphicx}
\expandafter\let\csname equation*\endcsname\relax 
\expandafter\let\csname endequation*\endcsname\relax 
\usepackage{amsmath}
\usepackage{color}
\usepackage{yfonts}

\usepackage{hyperref}
\hypersetup{
    colorlinks=true,
    linkcolor=blue,
    filecolor=magenta,
    urlcolor=cyan,
}
\newcommand{\Hess}{\mathrm{det}[\mathrm{d}^2H]}
\newcommand{\LV}{l}
\newcommand{\LVT}{l^2}

\newcommand{\grad}{\bm{\nabla}}

\newcommand{\IN}{\mathrm{L}}
\newcommand{\OU}{\mathrm{R}}
\renewcommand{\sp}{\mathit{sp}}

\newcommand{\tp}{\mathit{tp}}

\newcommand{\lr}{\mathit{lr}}
\newcommand{\ex}{\mathrm{ex}}

\begin{document}

\title{Quasinormal modes in Lorentz violating black hole analogues}

\author{Sam Patrick}
\email{samuel\_christian.patrick@kcl.ac.uk}
\affiliation{Department of Physics, King’s College London, University of London, Strand, London, WC2R 2LS, UK}

\author{Leonardo Solidoro}
\address{School of Mathematical Sciences, University of Nottingham, Nottingham, NG7 2RD, United Kingdom}

\date{\today}

\begin{abstract}
Analogue models of black holes typically have collective excitations with a dispersion relation that breaks the effective Lorentz symmetry at high energy. We investigate the consequences of such Lorentz violations on the quasinormal modes (QNMs) of the system, that is, the modes of energy dissipation. The model involves low frequency capillary-gravity waves (whose dispersion relation can be adjusted between sub- and superluminal modifications) around a draining vortex, which mimics a rotating black hole spacetime. For a subluminal/superluminal modification, the frequency and decay rate of co-rotating modes can be reduced/increased substantially, whilst counter-rotating modes are barely affected. A further consequence of the superluminal modification is that there are no co-rotating QNMs above a critical rotation and no counter-rotating ones below a critical rotation for strong enough Lorentz violations.
\end{abstract}

\maketitle


\section{Introduction}


Quasinormal modes (QNMs) are damped oscillations of open systems \cite{berti2009review}. Mathematically, they arise as complex eigenvalues $\omega_\mathrm{QNM}$ of non-conservative equations of motion, where $\mathrm{Re}[\omega_\mathrm{QNM}]$ is the oscillation frequency and $|\mathrm{Im}[\omega_\mathrm{QNM}]|^{-1}$ is the lifetime.
Provided the system does not suffer an instability, $\mathrm{Im}[\omega_\mathrm{QNM}]<0$ and QNMs decay in time as the system relaxes toward equilibrium.
Although they arise in many areas of physics (e.g. damped harmonic oscillators, nuclear resonances \cite{gamow1928quantentheorie}, leaky cavities \cite{leung1994completeness}, optical resonators \cite{kristensen2015normalization}), they are perhaps best associated with the ringdown of astrophysical black holes (BHs), having been detected in gravitational wave signals from binary BH mergers \cite{abbott2016observation}.
QNMs are of intense interest since their frequencies depend only on the system's properties and not the disturbance that stimulates them. This makes QNMs useful probes of systems which are difficult to access, such as black holes and neutron stars \cite{kokkotas1999stars,dreyer2004black}.
For astrophysical BHs in empty space (which are well described by the Kerr solution) the QNM spectrum is uniquely determined by the BH mass $M$ and rotation parameter $a$.
Realistic BHs are surrounded by matter, which has prompted research into environmental effects on the QNM spectrum \cite{leung1997quasinormal,barausse2014can,cannizzaro2024impact} and its stability under small perturbations \cite{jaramillo2021pseudospectrum,cheung2022destabilizing}.

Besides environmental effects, it is widely accepted that general relativity is not the final theory of gravity, hence, modifications to gravity may themselves imprint on the QNM spectrum.
A feature of some of these theories is that they are no longer Lorentz invariant (LI) at energies approaching the Planck scale.
Lorentz violating (LV) extensions are motivated by various quantum gravity inspired proposals, e.g. string theory, loop quantum gravity, doubly-special relativity, as well as modified gravity theories such as Ho\v{r}ava-Lifshitz and Einstein-Aether gravity (see e.g.~\cite{mattingly2005modern,amelino2002relativity,liberati2007quantum,liberati2013tests} and references therein).
In LV theories which preserve rotational and parity invariance, the energy-momentum ($E-P$) relation for massless elementary particles can be expanded at low momenta as,
\begin{equation} \label{E-M}
E^2 = c^2P^2 + \frac{c^2l^2P^4}{\hbar^2} + ... \ ,
\end{equation}
where the last term marks the departure from the LI theory and onsets below some length scale $l$.
In general, LV effects are expected influence the dynamics of matter around BHs, see e.g. \cite{barausse2011black,barausse2013black} for examples.

One scenario where LV effects in \eqref{E-M} arise naturally is in analogue models of BHs \cite{barcelo2011analogue}.
The idea (pioneered by Unruh \cite{unruh1981experimental} then rediscovered by Visser \cite{visser1993lorentzian}) is that long-wavelength excitations of an inhomogenous medium (often an accelerating fluid in the context of BHs \cite{unruh1981experimental}) obey an equation of motion which is the same that of a massless scalar field in a curved spacetime.
This equivalence at low energies implies not only that predictions of quantum field theory in curved spacetime can be tested in laboratory analogues, but also that the robustness of such effects to high energy modifications can be studied.
The low energy theory is LI, with all massless excitations propagating at a constant speed $c$, which is the analogue speed of light.
LV modifications at high energies lead to wave dispersion, that is, high-frequencies travel either slower or faster than $c$ depending on whether the $E-P$ relation, or equivalently the dispersion relation, is sub- or superluminal.

Along this line, early theoretical studies demonstrated the insensitivity of the Hawking effect to high energy LV modifications \cite{unruh1995sonic,corley1996spectrum} and later, both the stimulated and spontaneous Hawking effects were measured in laboratory experiments \cite{rousseaux2008observation,weinfurtner2011measurement,euve2016observation,steinhauer2016observation,munoz2019observation,kolobov2021observation}.
Rotating BH analogues exhibit additional phenomena, with superradiant amplification being measured in \cite{torres2017rotational,braidotti2022measurement} and its modification due to wave dispersion studied in \cite{patrick2020superradiance,patrick2021rotational,patrick2024primer}.
Characteristic excitations of these systems consistent with QNM oscillations were also measured in \cite{torres2020quasinormal,smaniotto2025black} in the regime where the system is strongly dispersive.
State-of-the-art simulators of rotating BH phenomenology employing superfluid $^4$He \cite{svanvcara2024rotating} and polariton fluids \cite{delhom2024entanglement} are now building toward measurements of these processes in regimes where excitations, and the effective spacetime they scatter with, exhibit quantum mechanical features, e.g. quantised angular momentum.
In these systems, one of the consequences of quantum mechanics is precisely an LV modification of the form \eqref{E-M} to the $E-P$ relation of low energy quasiparticles.
Hence, these systems are poised as a valuable tool (both experimental and theoretical) for investigating the interplay of quantum effects and high energy physics in curved spacetime scenarios.

In this work, we add to this on-going investigation by studying the influence of LV effects on the QNM spectrum of a specific analogue rotating BH model: surface waves around a draining vortex.
A key aim will be to analyse the effect of a perturbative LV modification to the QNM spectrum, in particular, the different signatures of sub- and superluminal modifications.
To calculate the QNM frequencies, we adapt the Wenzel-Kramers-Brillouin (WKB) method used in \cite{torres2020estimate,patrick2020superradiance,patrick2021rotational,patrick2024primer} to scattering scenarios for QNMs, generalising the usual WKB-QNM formula \cite{iyer1987black1} to the LV case.
Although QNMs in the same system were already analysed in \cite{torres2018waves}, the present study is different in two ways.
Firstly, \cite{torres2018waves} assumed at the outset that the QNMs can be treated as waves trapped on stationary circular orbits (a well-known approximation in the LI case \cite{cardoso2009geodesic}), whereas in our case, we will derive this through a proper treatment of extra spatial modes that arise as a consequence of wave dispersion.
Secondly, \cite{torres2018waves} considered the case of strong dispersion whereas we will be concerned with perturbative corrections of the form \eqref{E-M}, allowing us to characterise deviations from the LI results.

It is known that WKB methods for computing QNM frequencies are susceptible to errors in certain cases, especially when searching for rapidly damped modes and using higher-order methods \cite{konoplya2019higher,matyjasek2024accurate}. We will primarily be interested in the least damped modes (those being the most observationally relevant). Furthermore, we will show that the lowest order WKB result already shows good agreement with numerically precise results in the LI case where standard computational algorithms are available \cite{cardoso2004qnm,matyjasek2024accurate}.

The rest of the paper is organised as follows.
In Section~\ref{sec:system}, we introduce our model and the equations of motion governing linear perturbations.
In Section~\ref{sec:LI}, we recap the usual treatment in the LI case (e.g.~\cite{berti2004qnm}), stating the WKB-QNM condition which we seek to generalise to the LV case.
In Section~\ref{sec:WKB}, we review the WKB method of \cite{torres2018waves,patrick2020superradiance,patrick2021rotational,patrick2024primer} as a means of generalising the notion of scattering potentials and the stationary circular orbit (or light-ring) to the dispersive case.
In Section~\ref{sec:saddle}, we study wave scattering around a saddle point and derive the approximation of QNMs in terms of the light-ring frequency after implementing the correct boundary conditions.
In Section~\ref{sec:results}, we show how LV effects modify the QNM spectrum and interpret the results using the properties of the light-ring.
We conclude in Section~\ref{sec:conc} and discuss our results in the context of real black holes.

\section{The system} \label{sec:system}

Consider a draining vortex in the centre of a large water tank.
The floor of the container (located at $z=0$) is everywhere flat except in the centre where there is a small hole through which water is allowed to drain.
Water is resupplied at the edge of the container in such a way that the asymptotic water level at $z=h$ remains constant.
Relevant experimental set-ups can be found in \cite{torres2017rotational,torres2020quasinormal} and a schematic illustration in Fig.~1 of \cite{patrick2024primer}.
We consider the dynamics of the vortex in a region which is far enough from the drain hole that the water level is approximately spatially uniform whilst the boundary in the $(x,y)$ plane is far enough away that the system can be considered infinite (see \cite{solidoro2024quasinormal,patrick2024primer} for a discussion of finite size effects).
Assuming that the fluid is shallow, irrotational and axisymmetric about the drain hole, the velocity field is given by,
\begin{equation} \label{vortex}
    \mathbf{v} = v_r\hat{\mathbf{e}}_r + v_\theta\hat{\mathbf{e}}_\theta, \quad v_r = -D/r, \quad v_\theta = C/r,
\end{equation}
where the parameters $(D,C)$ are positive constants (drain and circulation respectively) which can be considered as the analogue of the Kerr BH parameters $(M,a)$.
Note, since we are assuming that $h$ is constant, the vertical component of the velocity field can be consistently set to zero.
Since the fluid is irrotational in this region, the velocity field can be written as $\mathbf{v}=\grad\Phi$, where $\Phi$ is called the velocity potential.

The dynamics of long wavelength surface waves are described by the action,
\begin{equation} \label{action0}
\begin{split}
    S = & \ \int dt d^2\mathbf{x}\Big(-\eta D_t\phi  + \frac{h}{2}|\grad\phi|^2 - \frac{h^3}{6}(\nabla^2\phi)^2 \\
    & \qquad \qquad \qquad \qquad -\frac{g}{2}\eta^2 -\frac{\sigma}{2\rho}|\grad\eta|^2 \Big),
\end{split}
\end{equation}
where $(\phi,\eta)$ are perturbations to $(\Phi,h)$ and \mbox{$D_t = \partial_t+\mathbf{v}\cdot\grad$} is the material (convective) derivative.
Here, $g$ is the gravitational acceleration, $\sigma$ is the surface tension and $\rho$ is the density of water.
The action \eqref{action0} can be obtained as an expansion of the full capillary-gravity wave action \cite{patrick2024primer} where only the terms involving up to four powers of $-ih\grad$ are retained. Note that since our treatment assumes constant $h$, there is no issue with the operator ordering.
Stationarity of the action under independent variations of $\phi$ and $\eta$ result in the equations of motion,
\begin{equation} \label{wave_eqn}
\begin{split}
    D_t\phi+g\eta-\frac{\sigma}{\rho}\nabla^2\eta = & \ 0, \\
    D_t\eta + h\nabla^2\phi + \frac{h^3}{3}\nabla^4\phi = & \ 0.
\end{split}
\end{equation}

\section{Lorentz-invariant regime} \label{sec:LI}

Here, we review the standard approach which leads to the analogy with field theory in curved spacetime \cite{visser1998acoustic}.
Dropping the last term in each equation in \eqref{wave_eqn} amounts to only keeping the leading term in a low $k$ expansion.
In that case, the two can be combined into a single equation of the form,
\begin{equation} \label{KGeq}
    \frac{1}{\sqrt{-g}}\partial_\mu\left(\sqrt{-g}g^{\mu\nu}\partial_\nu\phi\right) = 0.
\end{equation}
This is the Klein-Gordon equation that describes the propagation of a massless scalar field $\phi$ in a curved spacetime.
The metric $g_{\mu\nu}$, corresponding to the effective curved spacetime, is determined by the velocity field of the vortex, and has following line element,
\begin{equation} \label{PG}
    g_{\mu\nu}dx^\mu dx^\nu = -c^2 dt^2 + \left(dr+\frac{D}{r}dt\right)^2 + \left(r d\theta-\frac{C}{r} dt\right)^2,
\end{equation}
with $x^\mu=(t,\mathbf{x})$.
Since \eqref{KGeq} is invariant under Lorentz transformations, the effective theory for long wavelengths is LI, except in this case, it is the shallow water wave speed $c=\sqrt{gh}$ which adopts the role of the speed of light.
This metric has a horizon $r_h$ and an ergosphere $r_e$ at the locations,
\begin{equation}
    r_h = \frac{D}{c}, \qquad r_e = \frac{\sqrt{C^2+D^2}}{c},
\end{equation}
and, hence, can be identified with an analogue rotating black hole spacetime.

The usual analysis proceeds by casting \eqref{KGeq} in the form of a one-dimensional Schr\"odinger equation with an effective potential barrier whose form is determined by the curvature of the metric. 
First $\phi$ is written as,
\begin{equation} \label{expans2}
    \phi = e^{im\theta-i\omega t}\exp\left(-i\int\frac{v_r\tilde{\omega}(r)dr}{c^2-v_r^2}\right)\psi(r),
\end{equation}
where the symmetries of \eqref{vortex} have been used to decompose into the different $\omega$ (frequency) and $m$ (azimuthal) modes (in general \eqref{expans2} is a sum over $\omega,m$ modes but here we focus each mode individually for simplicity).
The frequency in a frame rotating with the vortex is given by,
\begin{equation}
\tilde{\omega}(r) = \omega-\frac{mC}{r^2}.
\end{equation}
Next, one defines the tortoise coordinate,
\begin{equation} \label{tortoise}
    r_* = \frac{r}{c} + \frac{r_h}{2c}\log\left|\frac{r-r_h}{r+r_h}\right|,
\end{equation}
which maps the horizon to negative infinity in $r_*$.
The result is that \eqref{KGeq} becomes,
\begin{equation} \label{1d_eq}
    \left[-\partial^2_{r_*}+V(r)\right](\sqrt{r}\psi) = 0,
\end{equation}
where the effective potential is,
\begin{equation} \label{potential}
    V = -\tilde{\omega}^2 + \left(c^2-\frac{D^2}{r^2}\right)\left(\frac{m^2-1/4}{r^2}+\frac{5D^2}{4c^2r^4}\right).
\end{equation}
Note that \eqref{1d_eq} with \eqref{potential} is formally equivalent to the radial equation for perturbations of a 5-dimensional Schwarzschild black hole for a specific choice of parameters \cite{matyjasek2021accurate,matyjasek2024accurate}.
We will eventually study the eikonal limit $|m|\gg 1$, in which case the last term in parentheses in \eqref{potential} can be taken as $m^2/r^2$.
The classical turning points correspond to the zeros of $V$, and their positions determine how much reflection occurs when the wave scatters with the barrier.
To find the turning points as a function of frequency, it is useful to write,
\begin{equation} \label{V_WKB}
\begin{split}
    V = & \ -(\omega-\omega^+)(\omega-\omega^-), \\ 
    \omega^\pm = & \ \frac{mC}{r^2} \pm \sqrt{\left(c^2-\frac{D^2}{r^2}\right)\frac{m^2}{r^2}}.
\end{split}
\end{equation}
The two curves $\omega^\pm$ can be understood as generalised potentials in the sense that when $\omega=\omega^+$ (or $\omega=\omega^-$) there will be a classical turning point (this is the equivalent condition to $E=V$ in the standard Schr\"odinger equation).
Hence, $\omega^\pm$ can be used to infer all the scattering information about the system (at least in the eikonal limit).
We make this identification here since, although the potential $V$ does not carry over to the dispersive system, the functions $\omega^\pm$ will.

The well-known WKB approximation to the QNM frequencies involves expanding the potential in the vicinity of its maximum $r_{*0}$, (which satisfies $V'_0\equiv \partial_{r_*}V|_{r=r_{*0}}=0$) and requiring the local solution to be asymptotically in-going toward the horizon and out-going toward spatial infinity \cite{iyer1987black1,berti2009review}. One then obtains a condition,
\begin{equation} \label{QNM0}
    \frac{V_0}{\sqrt{-2V''_0}} = i\left(n+\frac{1}{2}\right),
\end{equation}
which can be solved for the complex QNM frequencies $\omega_\mathrm{QNM}(n)$, where $n$ is called the overtone number.
Hence, the principle goal of this work will be to identify the dispersive analogue of \eqref{QNM0}, then study its solutions to understand LV effects on the QNM spectrum.

\section{WKB method} \label{sec:WKB}

To assess the impact of the additional terms in \eqref{wave_eqn}, we will apply the ray-tracing method of \cite{torres2017rotational} which can easily handle modified dispersion relations that break LI.
The approach is equivalent to the usual WKB method used to compute QNM frequencies (e.g. \cite{iyer1987black1}) except, rather than applying the method to the one-dimensional equation in \eqref{1d_eq}, we start directly from the full equations of motion \eqref{wave_eqn}.
We first discuss how generic wave scattering is described in this formalism before applying our framework to study QNMs (which can be viewed as a particular type of wave scattering).

If $\mathbf{v}$ is assumed to vary smoothly, the wave can be split into a slowly varying amplitude and rapidly varying phase,
\begin{equation} \label{WKBansatz}
    \begin{pmatrix}
        \phi \\ \eta
    \end{pmatrix} = \begin{bmatrix}
        A(\mathbf{x},t) \\ B(\mathbf{x},t)
    \end{bmatrix} e^{i\mathcal{S}(\mathbf{x},t)}.
\end{equation}
When derivatives act on the fields $(\phi,\eta)$, we assume that terms involving derivatives of the amplitude or second derivatives of the phase are much smaller than terms involving only single derivatives of the phase, i.e. \mbox{$|\partial\mathcal{S}|^2\gg|\partial^2\mathcal{S}|$}, \mbox{$|\partial\mathcal{S}|\gg |\partial\log A|$} and similarly for $B$.
This assumption then establishes a hierarchy of equations which can be solved order by order.
Defining the local wavevector and frequency by,
\begin{equation}
    \mathbf{k} = \grad\mathcal{S}, \qquad \omega = -\partial_t\mathcal{S},
\end{equation}
the leading order contribution from \eqref{wave_eqn} is the dispersion relation,
\begin{equation} \label{disp1}
    \Omega^2 = c^2(k^2+\LVT k^4) + \mathcal{O}(k^6), \qquad \Omega=\omega-\mathbf{v}\cdot\mathbf{k},
\end{equation}
where $\Omega$ is the frequency in a reference frame comoving with the fluid, and we have kept only the two leading terms in powers of $k=|\mathbf{k}|$ \footnote{Note that in a quantum description, \eqref{disp1} is related to \eqref{E-M} in the rest frame of the fluid by a factor $\hbar^2$, in which case it gives the $E-P$ relation for quantised collective excitations of the fluid}.
We have also defined the LV parameter,
\begin{equation} \label{Lambda}
   \LVT = \frac{\sigma}{g\rho} - \frac{h^2}{3},
\end{equation}
where $\sqrt{\sigma/g\rho}$ is the capillary length, which determines when surface tension becomes important, and $\LVT$ is either positive or negative depending on the relative size of the capillary length to the water height.
Note that $|\LV|$ is a lengthscale below which Lorentz invariance ceases to be valid, mimicking the Planck length where one expects new physics to arise.
The truncation of \eqref{disp1} to quartic order leads to spurious behaviour for certain frequencies when $\LVT<0$, since the right hand side of \eqref{disp1} has two maxima in $k$, whereas the full dispersion relation does not (see e.g. \cite{patrick2024primer}).
These spurious effects can be avoided provided one works in a restricted frequency range (see Appendix~\ref{app:restrict}).

To see that the term proportional to $\LVT$ is LV, we make contact with the LI equation in \eqref{KGeq} by writing \eqref{disp1} as,
\begin{equation} \label{dispLV}
    g^{\mu\nu} k_\mu k_\nu + \LVT k^4 + \mathcal{O}(k^6) = 0,
\end{equation}
where \mbox{$k^\mu = (-\omega,\mathbf{k})$} is the 4-wavevector.
The first term is what one would obtain from applying the WKB approximation to \eqref{KGeq} and is manifestly LI.
The second term breaks the Lorentz symmetry since it involves only the spatial part of the 4-wavevector.
Here we note that \eqref{dispLV} for $\LVT=0$ governs null geodesics on a curved spacetime, with $k_\mu$ related to the momentum of the particle at the spacetime point $x^\mu$.
With the LV modification, \eqref{dispLV} determines the trajectories (or rays) followed by high frequency wavefronts.
We can therefore think of these rays as paths traced by fictitious massless particles with dispersion relation \eqref{disp1}, which have been called hydrons \cite{purser1962water}.

\begin{figure} 
\centering
\includegraphics[width=\linewidth]{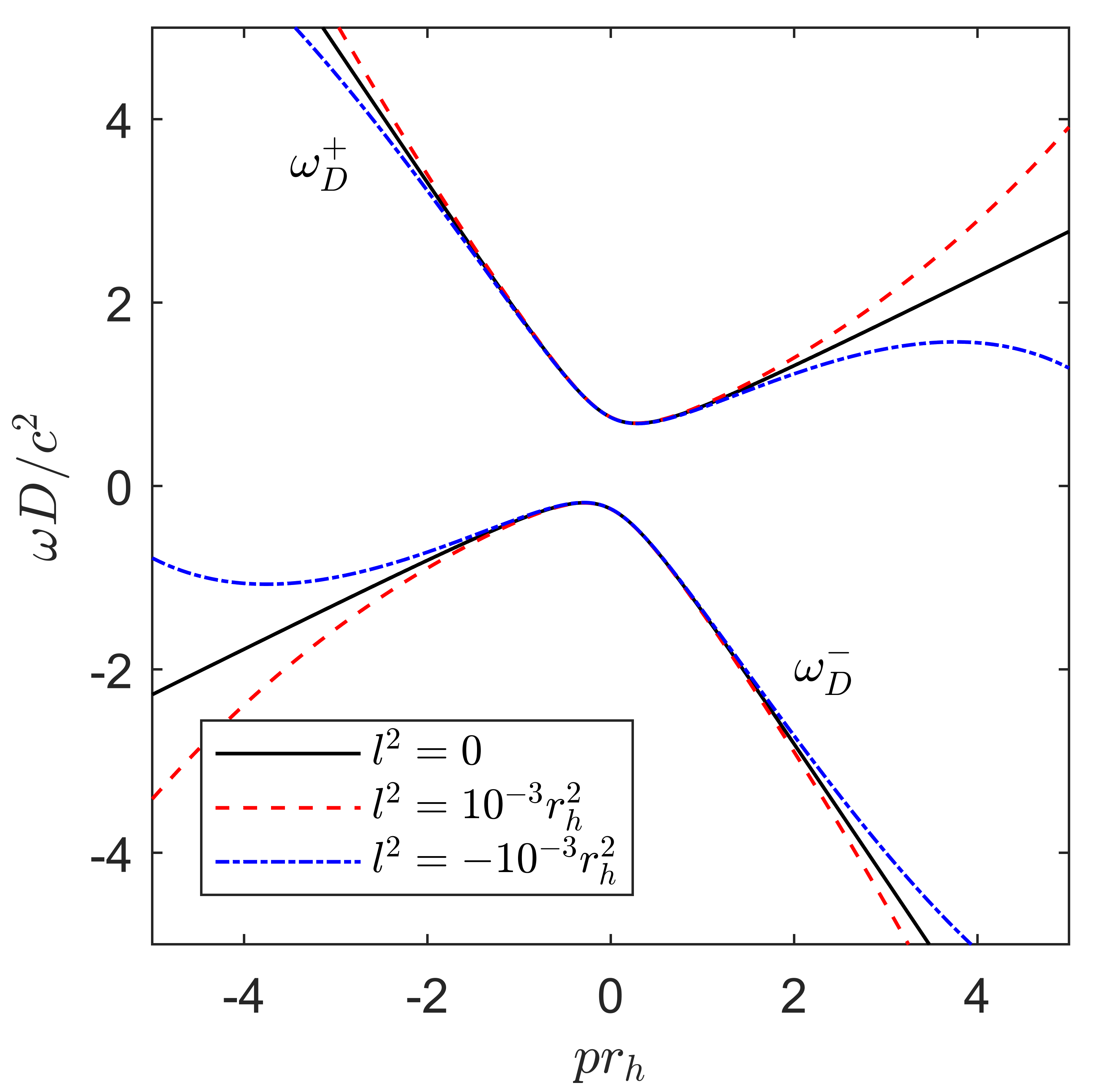}
\caption{The branches of the dispersion relation $\omega_D^\pm$ for the relativistic (LI) case $\LVT=0$ and the two dispersive (LV) cases with $\LVT\neq0$. The parameters in \eqref{disp1} are $C,D,c,m$ all $1$ and $r=2$. 
By fixing $m$, we are considering the part of the wave which propagates in the radial direction.
The gap at $p=0$ is due to the wave angular momentum (appearing as $m^2/r^2$ in the definition of $k$).
The tilt of the branches is a consequence of the radial velocity contribution to \eqref{disp1}, whilst the angular velocity has the effect of shifting both branches either up or down depending on the sign of $m$.
For $\LVT=0$, the branches are asymptotically linear in $p$ due the relativistic nature of this case, i.e. all frequencies travel at speed $c$.
For $\LVT>0$ ($\LVT<0$), the branches bend away from (toward) the $\omega=0$ axis due to the superluminal (subluminal) nature of the dispersion relation, i.e. higher frequencies travel faster (slower) than $c$.
} \label{fig:disp1}
\end{figure}

\begin{figure*} 
\centering
\includegraphics[width=\linewidth]{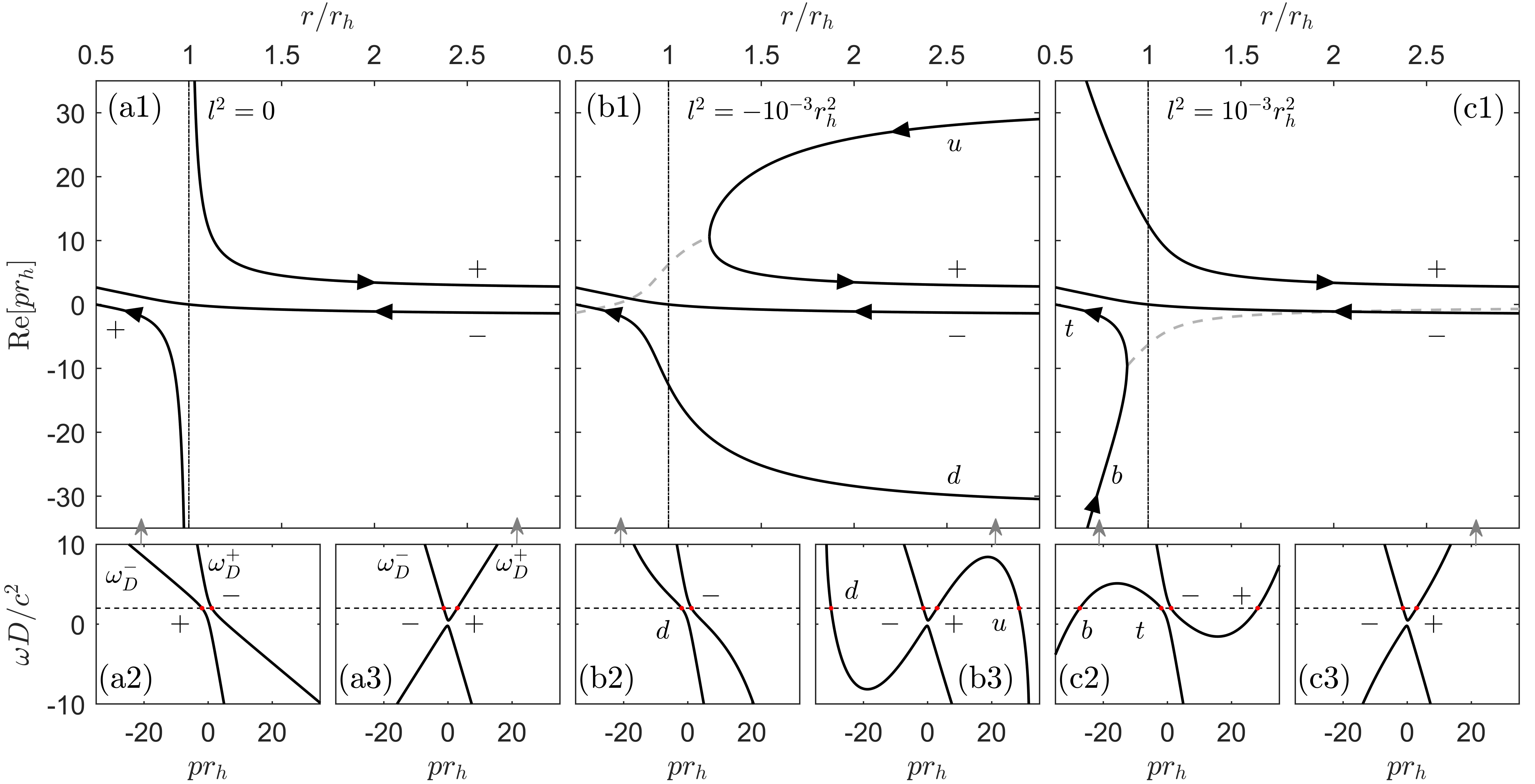}
\caption{Examples of the solutions $p^j$ to $\mathcal{H}=0$ as a function of radius. The parameters are $C/D,m$ both 1 and $\omega D/c^2=2$. The main panels show the ray trajectories (solid black lines) through $(r,p)$ phase space for the LI and two LV cases (only the real part of $p$ is shown). The dashed grey lines are where $p^j$ become evanescent. The horizon $r_h$ (only strictly defined when $\LVT=0$) is indicated as a vertical dashed-dotted line. In the LI regime [panels (a)], the out-going $p^+$ mode experiences an infinite blueshift at $r_h$. For $\LVT<0$ [panels (b)], $p^+$ gets blueshifted until dispersion converts it into an in-coming $u$ mode. For $\LVT>0$ [panels (c)], $p^+$ speeds up as it gets blueshifted and emanates from behind $r_h$.
The lower smaller panels illustrate the location of the $p^j$ on the two branches of the dispersion relation $\omega_D^\pm$ in the fast and slow flow regions ($r/r_h=0.75$ and $2.75$ respectively, indicated by grey arrows).
} \label{fig:rays}
\end{figure*}

More rigorously, rays are the worldlines of a collection of points distributed along wavefronts of the field $\phi$, that is, they are defined by the property that they trace out curves of constant phase. Hence, at the level of the eikonal approximation, i.e. ignoring amplitude effects in \eqref{WKBansatz}, the rays satisfy $\delta\mathcal{S}=0$, implying we can treat $\mathcal{S}$ as an effective ray action. To ensure that the dispersion relation \eqref{disp1} is satisfied, we add a constraint $\mathcal{H}=0$ to the action so that,
\begin{equation}
\mathcal{S} = \int d\lambda \left(k_\mu \dot{x}^\mu - \mathcal{H}\right),
\end{equation} 
where overdot is derivative with respect to $\lambda$, which parametrises the rays, and we have defined the effective Hamiltonian,
\begin{equation} \label{effH1}
\mathcal{H} = \frac{1}{2}g^{\mu\nu} k_\mu k_\nu + \frac{\LVT}{2} k^4
\end{equation}
Note that for $\LVT=0$, the corresponding Lagrangian is \mbox{$L=\frac{1}{2}g_{\mu\nu}\dot{x}^\mu\dot{x}^\nu$}, which is the typical one that governs geodesic motion.
In the Hamiltonian framework, minimising $\mathcal{S}$ results in Hamilton's equations,
\begin{equation} \label{Hamilton}
\dot{x}^\mu = \frac{\partial\mathcal{H}}{\partial k_\mu}, \qquad \dot{k}_\mu = -\frac{\partial\mathcal{H}}{\partial x_\mu}.
\end{equation}
These equations, subject to suitable initial conditions, can be solved for the ray trajectories, which govern how a wave scatters with the vortex.
There exists, however, a simpler approach in situations of high symmetry that involves reducing the dimensionality of the problem.
In particular, since $\mathbf{v}$ in \eqref{vortex} is independent of $\theta,t$, the decomposition into $\omega$ and $m$ modes is exact. 
Treating the problem in polar coordinates, the wavevector is $\mathbf{k}=(p,m/r)$ and the effective Hamiltonian can be expressed as,
\begin{equation} \label{effH2}
\begin{split}
\mathcal{H} = & \ -\frac{1}{2}(\omega-\omega_D^+)(\omega-\omega_D^-),
\end{split}
\end{equation}
where we have defined the functions,
\begin{equation} \label{disp_UL}
\begin{split}
\omega_D^\pm = & \ \frac{mC}{r^2} - \frac{pD}{r} \pm c\sqrt{k^2+\LVT k^4}, \\
k = & \ \sqrt{p^2+m^2/r^2}.
\end{split}
\end{equation}
Since \mbox{$(\omega,m,p)\to(-\omega,-m,-p)$} is a symmetry of \eqref{effH2}, we can restrict ourselves to $m\geq 0$ without losing any information.
In this case, $\omega>0$ modes co-rotate with the vortex whilst $\omega<0$ counter-rotate. 
When $\omega$ and $m$ are fixed, the remaining degrees of freedom are $p$ and $r$.
Now, rather than solving the remaining two equations in \eqref{Hamilton} for $p(\lambda)$ and $r(\lambda)$, we can instead view \eqref{effH2} as a function $\mathcal{H}(r,p)$. In our case, $\mathcal{H}$ is a quartic function of $p$ which we can solve for the roots $p^j(r)$ with $j=1...4$.
This fully solves the problem since the complete solution (for a given $\omega,m$ mode) can now be written as,
\begin{equation} \label{WKB_full}
\phi = \varphi(r) e^{im\theta-i\omega t}, \qquad  \varphi = \sum_j A^j(r)e^{i\int p^j(r) dr}.  
\end{equation}
where the amplitude $A^j(r)$ is uniquely determined by the trajectories $p^j(r)$ according to the amplitude equation.
We will not need to know the explicit form of this equation in this work (see e.g. \cite{patrick2020superradiance} where it is discussed in the context of superradiance).
A subtlety of Eq.~\eqref{WKB_full} is that the phase integral is evaluated relative to a point where the amplitude is known.
This allows one to relate WKB modes at two points $r_1$ and $r_2$ such that the WKB approximation is valid in the region $r\in[r_1,r_2]$.

We now focus on the ray trajectories $p^j(r)$. At each radius, we can think of the different $p^j$ as lying on one of the branches of the dispersion relation $\omega = \omega_D^\pm$ (see Fig.~\ref{fig:disp1}). The upper (lower) branch $\omega_D^+$ ($\omega_D^-$) is defined so that $\Omega>0$ ($\Omega<0$) which in the WKB approximation coincides with the sign of the wave norm density \cite{patrick2020superradiance}.
Hence $\omega_D^+$ and $\omega_D^-$ are the positive and negative norm branches respectively \footnote{Since energy is the product of the frequency and wave norm, negative energy solutions are allowed whenever $\omega>0$ are on the lower branch of the dispersion relation. This is the key mechanism underpinning superradiant scattering in rotating black holes and their analogues \cite{patrick2022quantum}.
In our discussion of quasinormal modes, however, this mechanism will not play a role.}. 
In Fig.~\ref{fig:rays}, we show some examples of the rays $p^j(r)$ for the LI and two LV cases, illustrating how these roots arise as solutions to $\omega=\omega_D^\pm$ at various radii.
In the LI case, there are two rays which we call these $p^\pm$ such that $\mathrm{Re}[p^+]>\mathrm{Re}[p^-]$.
For $\LVT<0$, the pair at low $p$ coincide with the LI ones and are also labelled $p^\pm$.
There are two additional modes which we label $p^{u,d}$ using the convention in \cite{patrick2020superradiance}, which satisfy $\mathrm{Re}[p^u]>\mathrm{Re}[p^+]$ and $\mathrm{Re}[p^-]>\mathrm{Re}[p^d]$.
For $\LVT>0$, there is also a pair which coincide with $p^\pm$, and the additional pair is labelled $p^{t,b}$, using the convention in \cite{patrick2021rotational}, such that $\mathrm{Re}[p^t]>\mathrm{Re}[p^b]$.

\begin{figure*} 
\centering
\includegraphics[width=\linewidth]{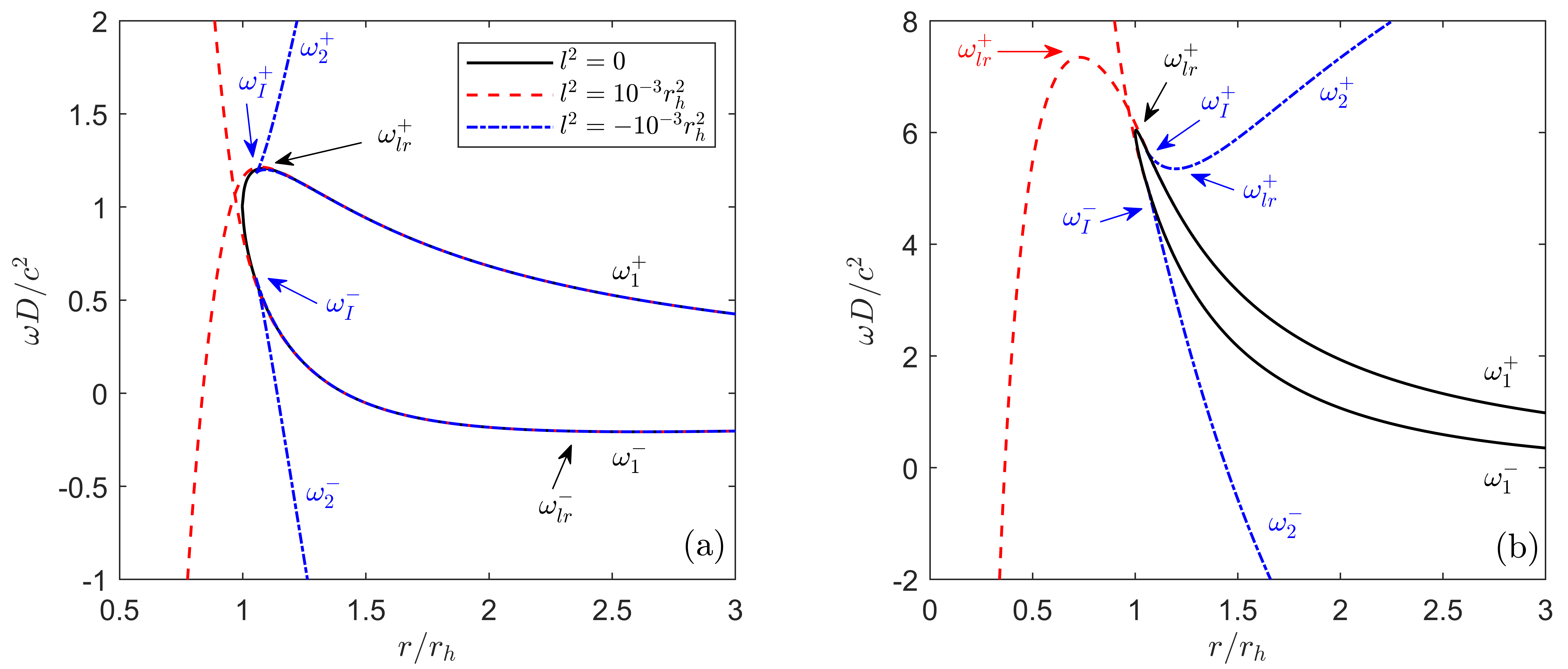}
\caption{The curves $\omega_a^\pm$ in Eq.~\eqref{WKB_pot} are plotted for various values of $\LVT$ and $C$.
For $\LVT=0$, these curves pinch off at the horizon, whereas for $\LVT>0$ they pass through the horizon.
For $\LVT<0$, the curves $\omega_1^\pm$ meet a second pair of curves $\omega^\pm_2$ at inflection points of $\mathcal{H}$ (labelled $\omega_I^\pm$).
The frequency of rays which orbit on the light-rings $r_\lr^\pm$ are labelled as $\omega_\lr^\pm$.
In panel (a), $C/D=1$ and $m=1$ and the two light-rings $r^\pm_\lr$ are on the curves labelled $\omega^\pm_1$ for all cases shown.
In panel (b), $C/D=6$ and $m=1$ and for $\LVT=-10^{-3}r_h^2$, $r_\lr^+$ is instead on the curve labelled $\omega_2^+$, which governs the scattering of the long wavelength out-going mode with a short-wavelength in-going one whose existence is a result of the sub-luminal wave dispersion.
} \label{fig:TP}
\end{figure*}

The direction of propagation (i.e. radially in or out) is determined by the radial component of the group velocity,
\begin{equation} \label{group}
    v^r_g \equiv \frac{dr}{dt} = -\frac{\partial_p\mathcal{H}}{\partial_\omega\mathcal{H}} = \partial_p\omega_D^\pm,
\end{equation}
where the first equality is a definition and we have used Hamilton's equations \eqref{Hamilton} and the form of $\mathcal{H}$ in \eqref{effH2} to deduce the second and third equalities.
We see that the direction of propagation is determined by the gradient of $\omega_D^\pm$ in the $p$ direction at the point $p^j$ where $\omega=\omega_D^\pm$ is satisfied.
Locations where $v_g^r=0$ are important since they correspond to locations where neighbouring rays intersect.
Physically, the wavefront travelling along the ray instantaneously comes to rest before reversing its direction.
Locations where this occurs are called turning points (corresponding to a point in phase space $(r_\tp,p_\tp)$) which, using Hamilton's equations \eqref{Hamilton}, can be found as the solutions to $\mathcal{H}=\partial_p\mathcal{H}=0$.
At turning points, the amplitude $A$ in \eqref{WKB_full} diverges and the WKB approximation breaks down.
Section~\ref{sec:saddle} will describe how to circumvent this failure of the method to describe the scattering of waves.
On the other side of a turning point, the two rays which have intersected are complex solutions of $\mathcal{H}=0$, corresponding to evanescent waves.
These solutions encode the ability of waves to tunnel through regions where they cannot propagate.

By \eqref{group}, we can use the gradient of the curves in Fig.~\ref{fig:rays} panels (a2,a3,b2,b3,c2,c3) to infer the direction of travel of the rays.
This information is added to the phase space plots in panels (a1,b1,c1) as arrows on the rays.
For $\LVT=0$, the out-going mode experiences an infinite blueshift (i.e. increase in $|\mathrm{Re}[p]|$) at the horizon, and for $r<r_h$ there are only in-going modes. In this region, the $+$ mode is on the negative norm branch.
For $\LVT<0$, the horizon is replaced by a turning point where the out-going $+$ mode gets blueshifted to higher $p$ then converted into an in-coming $u$ mode. This is enabled by the subluminal dispersion which reduces the group velocity of the ray at high $p$.
The negative norm solution labelled $d$ propagates everywhere and gets redshifted as it falls into the vortex.
For $\LVT>0$, the $+$ mode is again blueshifted but now its group velocity increases until it surpasses the flow speed, enabling it to emanate from inside the horizon.
The negative norm solutions labelled $t,b$ only propagate in the region of high radial velocity.

To study the scattering of waves, we now turn to a discussion of the turning points defined by $v_g^r=0$.
Using \eqref{group}, we see that the turning points correspond to the extrema of $\omega_D^\pm$ in the $p$ direction.
Let $p^\ex_a(r)$ be the location of these extrema on the $p$ axis, where $a$ indexes the number of extrema.
The value of $\omega_D^\pm$ at the extrema are then,
\begin{equation} \label{WKB_pot}
\omega^\pm_a(r) = \omega^\pm_D[r,p^\ex_a(r)],
\end{equation}
where the superscript on $\omega^\pm_a$ indicates whether the extremum is on the upper or lower branch of the dispersion relation.
These functions provide useful information since they determine the locations of the turning points through \mbox{$\omega=\omega^\pm_a(r_\tp)$}.
We display examples of these curves in Fig.~\ref{fig:TP}.
For $\LVT\geq0$, $\omega_D^\pm$ have a single real extremum per branch, hence, $a=1$.
For $\LVT<0$, there are two physical extrema on each branch ($a=1,2$).
The third extremum on each branch is unphysical and results from the truncation of the full dispersion relation to quartic order, hence, we do not consider it in the following.
Indeed, effects resulting from this feature of the dispersion relation can be avoided by working in the regime of validity stated in Appendix~\ref{app:restrict}.

The different $\omega^\pm_a$ correspond to the scattering of different WKB modes, or equivalently, the intersection of different pairs of neighbouring rays in $(r,p)$ phase space.
This is because at each $p$-extremum of $\omega_D^\pm$, two of the $p^j$ coalesce at the turning point.
For $\LVT=0$, both $\omega^\pm_1$ tell us when the rays labelled $\pm$ intersect.
Similarly for $\LVT>0$, at large $r$ both $\omega^\pm_1$ govern the intersection of $\pm$ rays whilst at low $r$ (where the radial velocity is high) $\omega^\pm_1$ can describe the intersection of $t,b$ rays.
This process is described in detailed in \cite{patrick2021rotational}.
For $\LVT<0$, $\omega^\pm_1$ governs the $\pm$ intersection, $\omega^+_2$ the $+,u$ intersection and $\omega^-_2$ the $-,d$ intersection.
The $r$ location where the $\omega^\pm_2$ meet the $\omega^\pm_1$ corresponds to an inflection point of the dispersion relation, since it corresponds to the coalescence of two $p$-extrema. At this location, $\mathcal{H}$ and its first two $p$-derivatives vanish which can be solved for the inflection point triplet $(r_I,p_I,\omega_I)$, see \cite{patrick2020superradiance,patrick2024primer} for details. The location of these inflection points is indicated on Fig.~\ref{fig:TP} with the labels $\omega_I^\pm$.

The key insight is the following. The QNM condition in \eqref{QNM0} refers to the properties of the effective potential $V(r)$, whose utility lies in the fact that its zeros tell us where to find the dominant contributions to scattering.
The issue with this formulation is that the notion of a scattering potential does not extend naturally to the dispersive regime since it relies on writing the equation of motion in the form \eqref{1d_eq}, whereas for $\LVT\neq 0$ the equations are higher (fourth) order in spatial derivatives.
However, the curves defined in \eqref{WKB_pot} are precisely those in \eqref{V_WKB} when $\LVT=0$.
Hence, \eqref{WKB_pot} are the relevant objects to study to determine how to extend the QNM condition \eqref{QNM0} to the dispersive regime.

For $\LVT=0$, QNMs involve frequencies that scatter near the maximum of $V(r)$.
In that case, the potential is expanded in the vicinity of its maximum and used to find an exact solution to the local wave equation.
In the dispersive case, we should instead expand $\mathcal{H}$ for frequencies which are close to the extrema of the $\omega^\pm_a$.
These extrema generalise the notion of a light-ring (or photon-sphere) in gravity, that is, a stationary circular orbit surrounding a black hole or compact gravitational object \cite{cardoso2009geodesic,torres2018waves}.
In terms of the effective Hamiltonian, each extremum of $\omega^\pm_a(r)$ corresponds to a triplet $(r_\lr,p_\lr,\omega_\lr)$ which solves,
\begin{equation} \label{LRcond}
\mathcal{H}_\lr = 0, \quad \partial_p\mathcal{H}_\lr=0, \quad \partial_r\mathcal{H}_\lr = 0.
\end{equation} 
These are stationary circular orbits, since Hamilton's equations \eqref{Hamilton} imply $\dot{r}_\lr =0$ and $\dot{p}_\lr =0$, i.e. this ray neither moves nor accelerates in the radial direction.
The equations \eqref{LRcond} can be solved to find the $r$ location of the light-ring,
\begin{equation} \label{rLR}
\begin{split}
    r^\pm_\lr =  & r_h\sqrt{\frac{B_\pm^2+1}{2B_\pm^2}\left(\beta_\pm+\beta_\pm^{1/2}\right)} \\
    B_\pm = & \ \frac{C\pm\sqrt{C^2+D^2}}{D}, \quad \beta_\pm = 1-4\gamma B_\pm^2,
    \end{split}
\end{equation}
where the subscript $\pm$ denotes the upper or lower branch of the dispersion relation.
We have also defined the dimensionless ratio,
\begin{equation}
\gamma = m^2l^2/r_h^2,
\end{equation}
which controls the strength of the LV modifications.
The light-ring momentum and frequency are given by \mbox{$p^\pm_\lr = m B_\pm/r^\pm_\lr$} and \mbox{$\omega^\pm_\lr = \omega_D^\pm(r^\pm_\lr,p^\pm_\lr)$} respectively.
In particular, when $|\gamma|\ll 1$ we have,
\begin{equation} \label{LR_lin}
\begin{split}
r^\pm_\lr = & \ r^\pm_{\lr 0}\left(1-3\gamma B_\pm^2/2 \right) +\mathcal{O}\left(\gamma^2\right), \\
\omega^\pm_\lr = & \ \omega^\pm_{\lr 0} \left(1+\gamma B_\pm^2\right) +\mathcal{O}\left(\gamma^2\right), \\
\omega^\pm_{\lr0} = & \ \frac{mc^2 B_\pm}{2D}, \qquad r^\pm_{\lr0} = r_h\left(\frac{2\sqrt{C^2+D^2}}{\pm B_\pm D}\right)^\frac{1}{2},
\end{split}
\end{equation}
where the subscript 0 indicates the values for $\LVT=0$.

\begin{figure} 
\centering
\includegraphics[width=\linewidth]{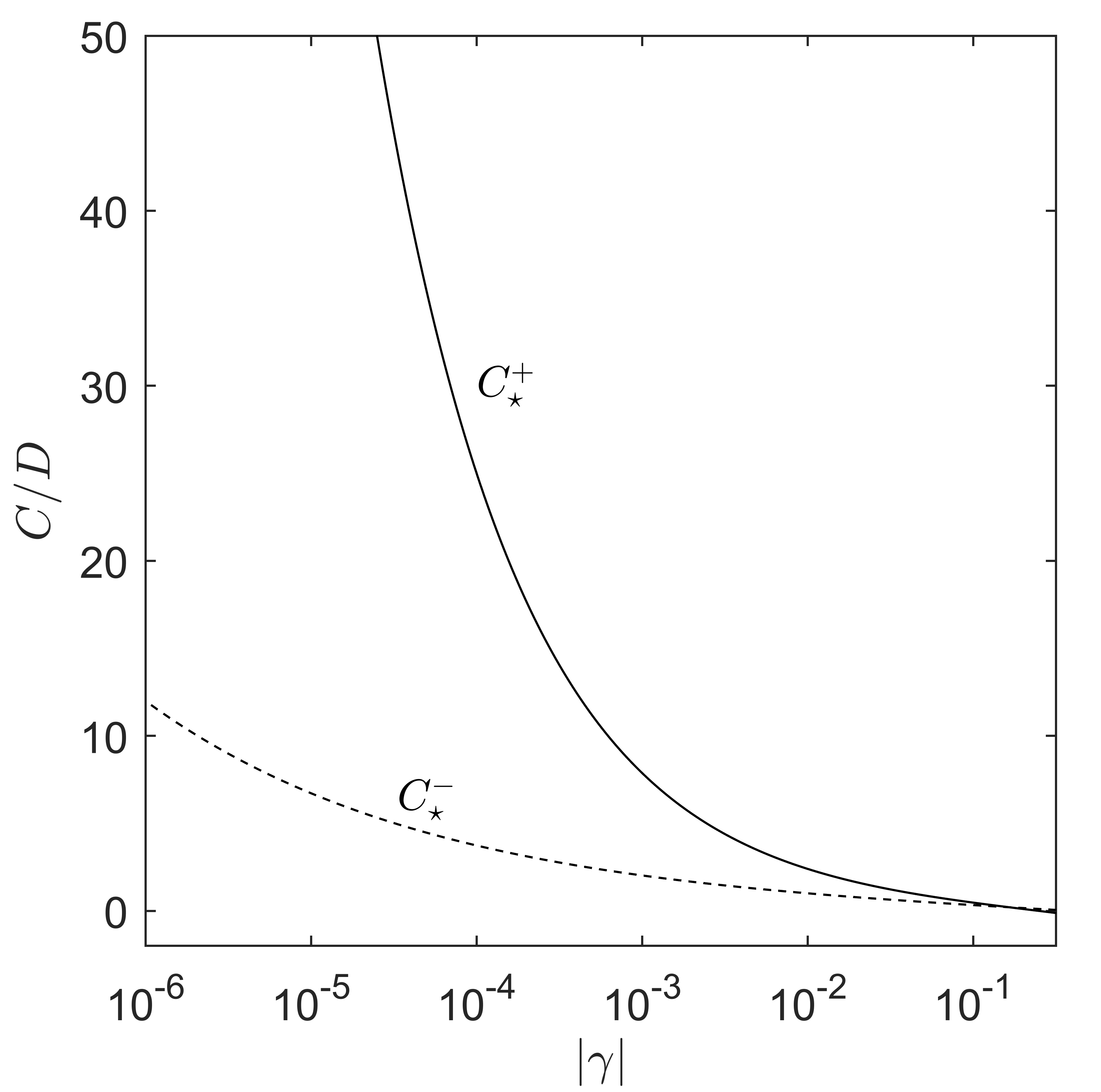}
\caption{Critical rotations as a function of the parameter $\gamma$ which controls the strength of LV modifications.
$C^-_\star$ (black dashed line) determines whether the co-rotating light-ring scatters $+,-$ or $u,+$ rays for $\gamma<0$, whilst $C^+_\star$ (solid black line) gives the rotation where the light-ring radius goes to zero for $\gamma>0$. 
} \label{fig:critC}
\end{figure}

\begin{figure} 
\centering
\includegraphics[width=\linewidth]{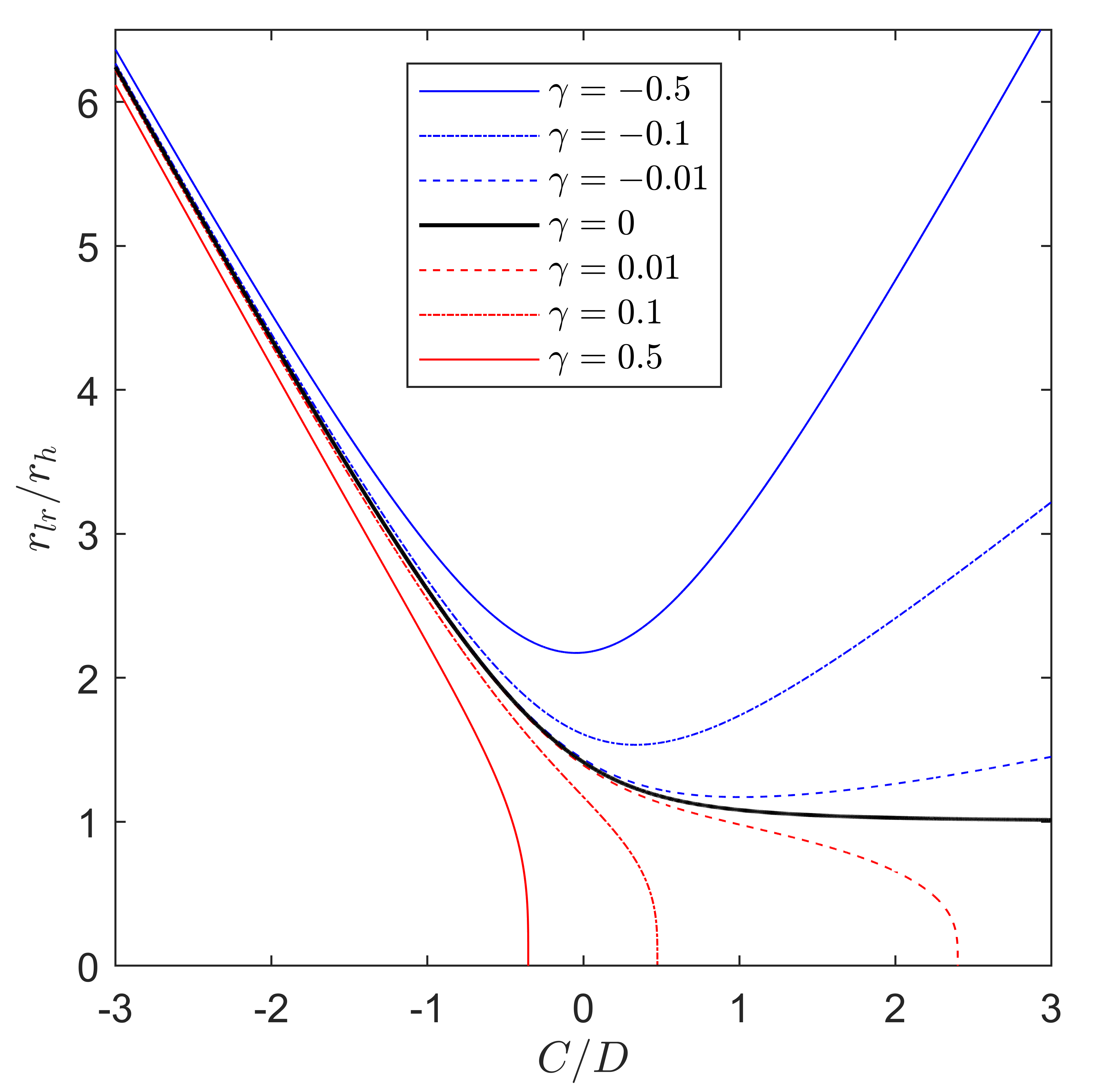}
\caption{The light-ring radius is plotted as a function of $C/D$ for several values of the LV strength $\gamma$.
The co- and counter-rotating light-rings $r_\lr^+$ and $r_\lr^-$ are shown for $C>0$ and $C<0$ respectively.
Computationally, $r_\lr^-$ is computed using Eq.~\eqref{rLR} assuming $(-\omega,m,C)$ are all positive numbers, before mapping the results to $(\omega,m,-C)$ using symmetries of the system.
} \label{fig:LR}
\end{figure}

The $\omega^\pm_\lr$ are indicated on the $\omega^\pm_a$ curves in Fig.~\ref{fig:TP} with labels.
An important observation is that for $\LVT<0$, $\omega_\lr^+$ can be on either $\omega^+_1$ or $\omega^+_2$, in which case the modes emanating from the light-ring are the $\pm$ pair or the $+,u$ pair respectively.
The former (latter) occurs when \mbox{$\omega^+_I<\omega_\lr^+$} \mbox{($\omega^+_I>\omega_\lr^+$)}, as shown in Fig.~\ref{fig:TP}.
This has important consequences for co-rotating QNMs with subluminal dispersion as we shall see later.
The transition between the two cases occurs when the light-ring frequency is an inflection point of the dispersion relation, that is when $\omega_\lr^+=\omega_I^+$.
This relation can be solved to find a critical $C^-_\star(\gamma)$ which varies with the strength of LV modifications, such that the co-rotating light-ring scatters the $+,-$ ($u,+$) rays for $C<C^-_\star$ ($C>C^-_\star$).
This critical value is shown in Fig.~\ref{fig:critC} and is inversely proportional to $|\gamma|$, that is, the light-ring scatters $(u,+)$ rays at lower rotations if LV effects are strong.

Another important consequence of the expression in \eqref{rLR} is that, for $\LVT>0$, there is a critical rotation at which the location of the light-ring goes to zero, which occurs when $\beta_\pm = 0$.
Solving this condition gives the critical value \cite{patrick2021rotational},
\begin{equation} \label{Cmax}
C^+_\star = D \frac{r_h^2-4m^2l^2}{4r_h m|l|},
\end{equation}
which is also plotted on Fig.~\ref{fig:critC}.
To appreciate the effect of this value, it is useful to plot the co- and counter-rotating light-ring radii $r_\lr^+$ and $r_\lr^-$ as a function of $C$, using the fact that \eqref{effH2} is symmetric under \mbox{$(-\omega,m,C,-p)\to(\omega,m,-C,p)$} to represent counter-rotating modes as positive frequencies with negative rotation.
This is the approach taken in Fig.~\ref{fig:LR}, where the co- and counter-rotating light-ring radii are shown for $C>0$ and $C<0$ respectively.
We see that in the LI case, the light-ring of counter-rotating modes moves further outside the vortex core for faster rotations, whereas the co-rotating light-ring gets close to the horizon.
For $l^2<0$ ($l^2>0$), the light-ring moves toward larger (smaller) radii relative to the LI location, with the effect being most pronounced for co-rotating modes.
When $\gamma>0$, the light-ring radius goes to zero at the rotation determined by \eqref{Cmax}.
For weak dispersion ($\gamma<1/2$), this means that there is no light-ring for co-rotating modes at high circulation (i.e. for $C>C^+_\star$) whereas the light-ring persists for all counter-rotating modes.
For strong dispersion ($\gamma>1/2$), there is no light-ring for any co-rotating modes and the light-ring is also absent for counter-rotating modes at low rotation ($|C|<|C^+_\star|$).
The existence of the critical rotations $C^\pm_\star$ underpin the features of the LV modified QNM spectrum we present in Section~\ref{sec:results}.

\begin{figure*} 
\centering
\includegraphics[width=.95\linewidth]{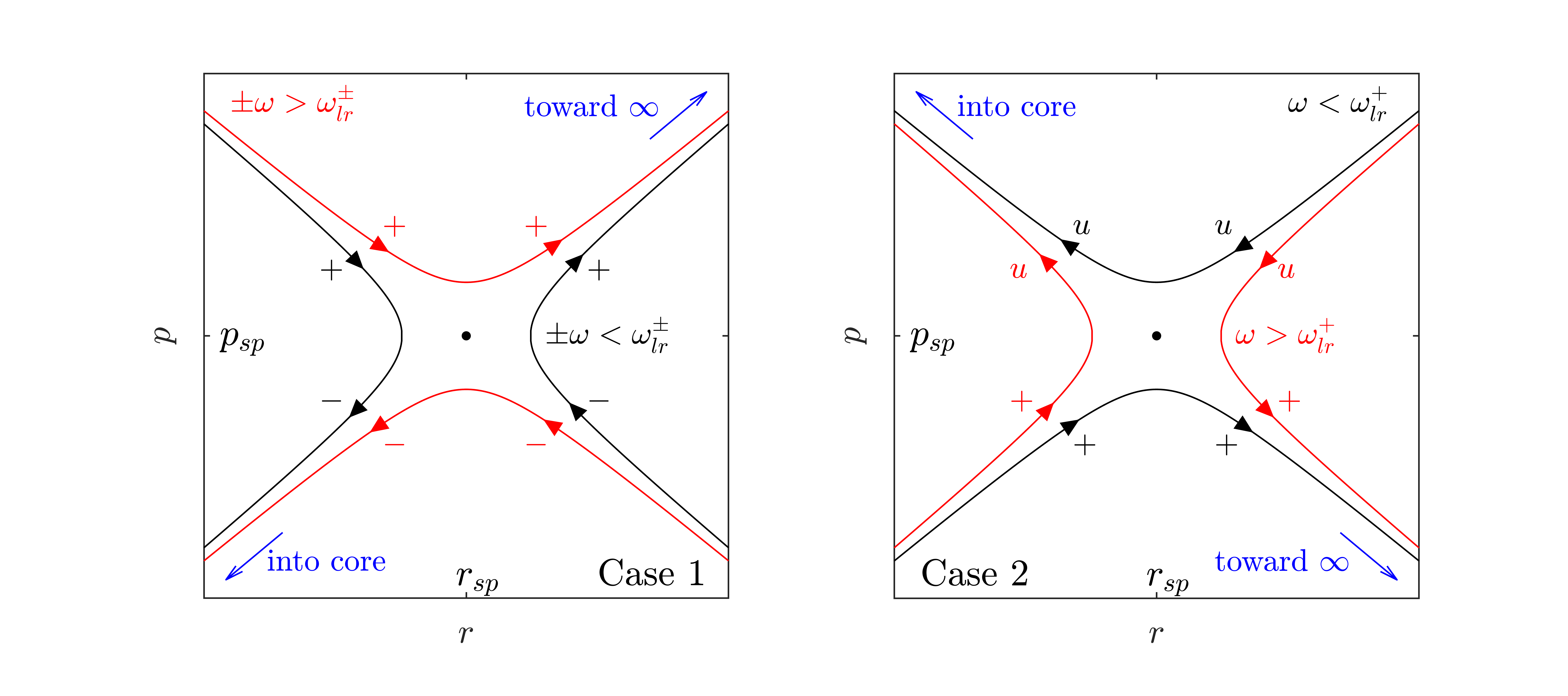}
\caption{Schematic ray trajectories in $(r,p)$ phase space which play a key role in QNMs. Scattering near the saddle point around a draining vortex either involves two long wavelength modes $p^\pm$ (case 1) or a long-wavelength out-going mode $p^+$ and a short wavelength in-going mode $p^u$.
In case 1, the rays are completely reflected by the saddle point at low frequency (black lines) whereas the rays glance the saddle point at high frequencies (red lines).
Fixing $m>0$, case 2 occurs only for $\omega>0$, and involves a reflected ray at high frequency (red line) and glancing rays at low frequency (black lines).
By studying the full scattering diagrams in \cite{patrick2020superradiance,patrick2021rotational}, we find that WKB waves moving away from the saddle point propagate all the way into the asymptotic region, allowing us to impose the QNM boundary conditions a small distance either side of the saddle point.
} \label{fig:saddle}
\end{figure*}

\section{Saddle point} \label{sec:saddle}

For frequencies close to $\omega_\lr$, we can expand $\mathcal{H}$ in the vicinity of a special point in phase space called the saddle point $(r_\sp,p_\sp)$.
This doublet is the solution to the equations,
\begin{equation} \label{sp_def}
\partial_p\mathcal{H}_\sp=0, \quad \partial_r\mathcal{H}_\sp = 0.
\end{equation}
which are solved for,
\begin{equation}
r^\pm_\sp = r_h\sqrt{\frac{C/D-B_\pm\left[1-2\gamma(B_\pm^2+1)\right]}{\omega D/mc^2-B_\pm}},
\end{equation}
and $p^\pm_\sp = B_\pm/r^\pm_\sp$ with $B_\pm$ given in Eq.~\eqref{rLR} (again, the $\pm$ denotes the upper or lower branch of the dispersion relation).
This point coincides with $r^\pm_\lr$ when $\mathcal{H}=0$ in the same way that the maximum of $V$ coincides with that of $\omega^\pm$ when $V=0$, and in both cases this occurs for $\omega=\omega_\lr$.
The benefit of expanding near the saddle point is that we need only consider the modes which are locally interacting, i.e. neighbouring rays that either intersect or come close to intersecting.

As described in Appendix \ref{app:saddle}, the effective Hamiltonian can be expanded to quadratic order in $r-r_\sp$ and $p-p_\sp$.
Restoring radial derivatives by setting $p\to -i\partial_r$, one can find an exact solution to the local wave equation which governs the scattering of the mode pair (either $p^\pm$ or $p^{+,u}$) interacting at the saddle point.
The asymptotic behaviour of this solution can then be matched onto the WKB modes to find a transfer matrix relating the WKB amplitudes either side of the saddle point,
\begin{equation} \label{RL_transfer}
\begin{split}
& \begin{pmatrix}
A^\OU_< \\ A^\IN_<
\end{pmatrix} = \frac{e^{-2i\chi}}{\tau}\begin{pmatrix}
\beta \alpha^{-2} & i \\ -i & \beta^*\alpha^2
\end{pmatrix} \begin{pmatrix}
A^\OU_>\\ A^\IN_>
\end{pmatrix}, \\
& \tau = e^{-\pi b}, \quad \beta = \frac{\sqrt{2\pi i\tau}}{\Gamma(\frac{1}{2}+ib)}, \quad \alpha = |b|^{-\frac{ib}{2}}e^{\frac{ib}{2}+\frac{i\pi}{8}}, \\
& \ \chi = \mathrm{max}(0,b)p_\sp(2\mathcal{H}/\partial^2_p\mathcal{H})^{1/2}_\sp,
\end{split}
\end{equation}
where the properties of the saddle point enter through the parameter,
\begin{equation}
b = \frac{\epsilon\mathcal{H}}{\sqrt{-\mathrm{det}[\mathrm{d}^2\mathcal{H}]}}\bigg|_\sp, \quad \mathrm{d}^2\mathcal{H} = \begin{pmatrix}
\partial^2_p\mathcal{H} & \partial_p\partial_r\mathcal{H} \\ \partial_p\partial_r\mathcal{H} & \partial^2_r\mathcal{H}
\end{pmatrix}
\end{equation}
with $\mathrm{d}^2\mathcal{H}$ the Hessian matrix. 
The parameter $\epsilon$ takes the values $1$ resp. $-1$ for the cases 1 resp. 2 in Fig.~\ref{fig:saddle}.
$A^j_i$ is the amplitude of the $p^j$ mode at the location $r_i$, and $\OU,\IN$ are a mode pair label such that \mbox{$\mathrm{Re}[p^\OU]>\mathrm{Re}[p^\IN]$} (with $\OU=+$ and $\IN=-$ for case 1 and $\OU=u$ and $\IN=+$ for case 2).
The locations $r_<$ and $r_>$ are turning points either side of $r_\sp$ when $b>0$, and both equal to $\mathrm{Re}[r_\sp]$ when $b<0$.
Note that $\omega$ is assumed real in the derivation of \eqref{RL_transfer} so that $b\in\mathbb{R}$.

The defining property of black hole QNMs is that they are in-going on the horizon and out-going at spatial infinity.
This is simple to implement in the LI case ($\LVT=0$) case since there are only two spatial modes and one can easily distinguish which is in-going and which is out-going.
In the LV ($\LVT\neq 0$) case, the situation is more complicated since there are more spatial modes (in our case four) corresponding to the different possible rays $p^j(r)$.
One must therefore take care to identify which ones are in-going and out-going toward $r=0$ and $r\to\infty$ respectively.

To do this, one can plot the phase space diagrams (like the ones shown in Fig.~\ref{fig:rays}) for the $\omega$ range of interest 
Although the precise phase space plots will differ depending on the choice of parameters $(m,C,D,c,\LVT)$, the different possibilities can be grouped into categories which share the same order of turning points and asymptotic mode properties.
This is the approach taken in \cite{patrick2020superradiance,patrick2021rotational}, where the different categories are referred to as scattering \emph{Types}, followed by a Roman numeral.
Ref.~\cite{patrick2021rotational} analyses the cases $\LVT\geq0$.
Ref.~\cite{patrick2020superradiance} studies the scattering of deep water ($hk\gg 1$) waves with $\sigma=0$, which share the same scattering classification as the $\LVT<0$ case considered here, provided we work within the constraints of Appendix~\ref{app:restrict} (this is due to the fact that both these cases are subluminally dispersive).
Rather than reproducing the full analyses here, we take from the results of \cite{patrick2020superradiance,patrick2021rotational} only those rays trajectories which are relevant for wave scattering near a saddle point.
There are two cases of interest corresponding to cases 1 and 2 on Fig.~\ref{fig:saddle}.

\textbf{Case 1:} the WKB modes which interact near the saddle point correspond to the $p^\pm$ rays. 
This is the only possibility when $\LVT\geq 0$.
For $\LVT<0$ this case occurs when $\omega_\lr^+>\omega_I^+$.
The relevant frequencies are those near $\omega_\lr^\pm$, which correspond to the scattering diagrams in \cite{patrick2021rotational} labelled Types I$^0$ and II$^0$ for $\LVT=0$ and Types I$^+$, II$^+$, V$^+$, and VIII$^+$ for $\LVT>0$.
In \cite{patrick2020superradiance}, the relevant diagrams are Types I and III.
One finds from inspecting these diagrams that the $p^+$ mode is the out-going one over the full $r$ range whereas the $p^-$ mode is in-going and the QNM BCs read,
\begin{equation} \label{QNMBC1}
\text{Case 1:}~\varphi \sim \begin{cases}
A^+_L e^{i\int^{r}_{L}p^+(r')dr'}, \quad r\to L \\
A^-_s e^{i\int^{r}_{s}p^-(r')dr'}, \quad r\to s
\end{cases}
\end{equation}
where $r=s$ ($r=L$) is a location in the small (large) $r$ region for which no further scattering takes place for $r<s$ ($r>L$).
These locations generalise $r=r_h$ and $r\to\infty$ where the BCs are usually applied.

\begin{figure*} 
\centering
\includegraphics[width=\linewidth]{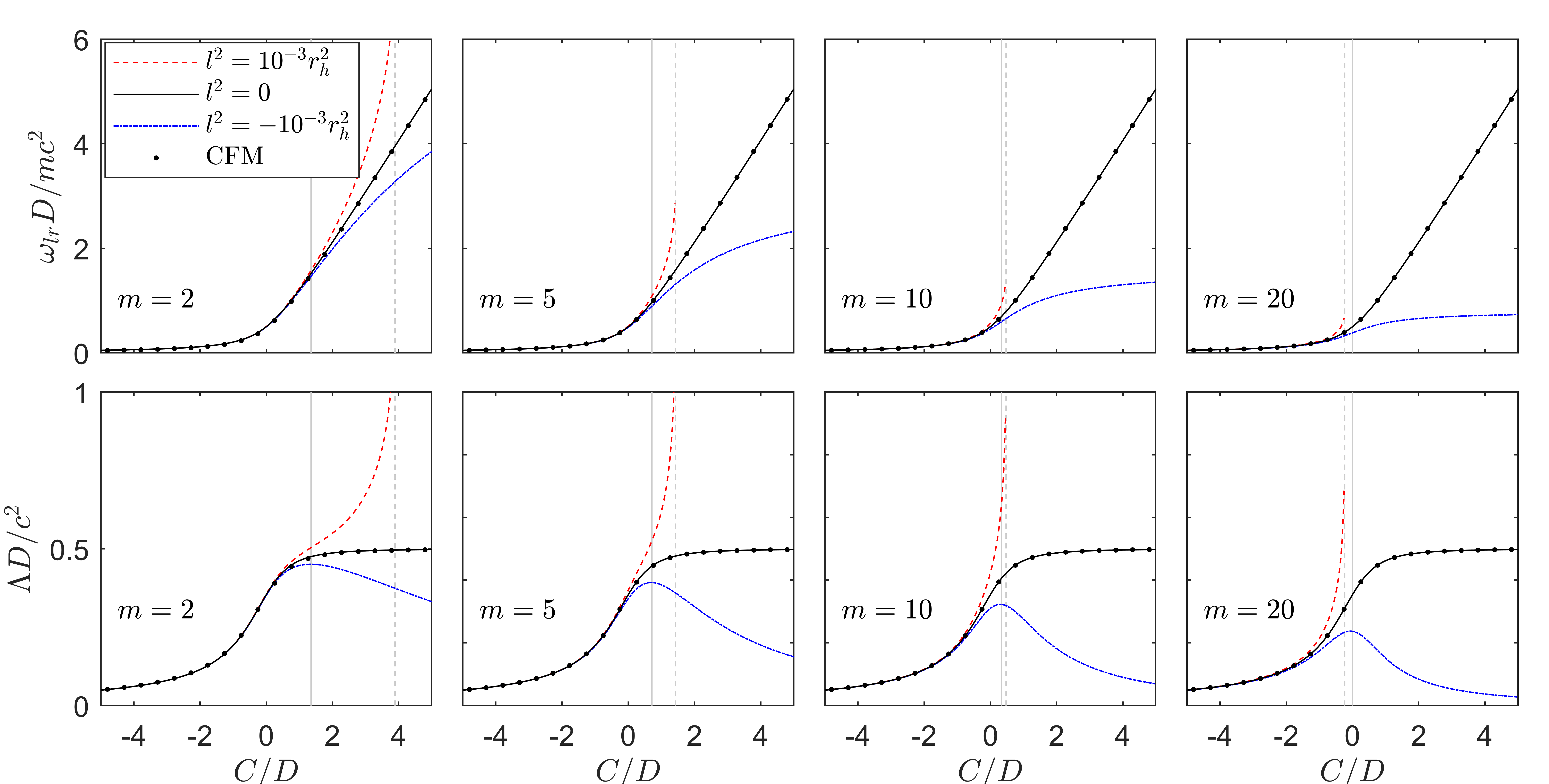}
\caption{The QNM spectrum in the eikonal limit is characterised by the light-ring frequency $\omega_\lr$ and Lypunov exponnent $\Lambda$ in Eq.~\eqref{qnm_final}. We display co- and counter-rotating QNMs for $C>0$ and $C<0$ respectively.
The LI case ($l^2=0$, black line) is compared against representative superluminal ($l^2>0$, red dashed line) and subluminal ($l^2<0$, blue dashed-dotted line) results for various $m$-modes, as well as exact results obtained using the continued fraction method (CFM, black dots) of \cite{cardoso2004qnm}.
The spectrum deviates from the LI one at critical rotations $C^-_\star$ (for $l^2<0$) and $C^+_\star$ (for $l^2>0$), which are marked by solid and dashed grey lines respectively.
These deviations are more pronounced for higher $m$-modes.
} \label{fig:Cdep}
\end{figure*}

\textbf{Case 2:} the WKB modes which interact near the saddle point correspond to the $p^{+,u}$ rays. 
This occurs for $\LVT<0$ when $\omega_\lr^+<\omega_I^+$. The relevant diagrams in \cite{patrick2020superradiance} are Types II, IV and VI.
Inspecting these diagrams shows that the $p^+$ mode is out-going in the large $r$ region whilst $p^u$ mode is in-going toward the centre, and the QNM BCs read,
\begin{equation} \label{QNMBC2}
\text{Case 2:}~\varphi \sim \begin{cases}
A^+_L e^{i\int^{r}_{L}p^+(r')dr'}, \quad & r\to L \\
A^u_s e^{i\int^{r}_{s}p^u(r')dr'}, \quad & r\to s
\end{cases}
\end{equation}
In the WKB approximation, the mode amplitudes at \mbox{$r=L$} \mbox{($r=s$)} are adiabatically connected to those at $r_>$ ($r_<$), that is, there there is no energy transfer as the modes traverse these regions and we are assuming that all scattering takes place at the saddle point.
Hence, \eqref{QNMBC1} and \eqref{QNMBC2} imply we should set $A^+_<=A^-_>=0$ in case 1 and $A^+_<=A^u_>=0$ in case 2.
Applying these conditions to the transfer matrix in \eqref{RL_transfer} (and using appropriate mode pair to stand in for $\OU,\IN$) we find that the QNM frequencies correspond to the poles of $\Gamma(\frac{1}{2}+ib)$ in case 1 and $\Gamma(\frac{1}{2}-ib)$ in case 2.
Using the definition of $b$, this means we can write the QNM condition in both cases as,
\begin{equation} \label{QNM1}
\frac{\mathcal{H}}{\sqrt{-\mathrm{det}[\mathrm{d}^2\mathcal{H}]}}\bigg|_\sp = i\left(n+\frac{1}{2}\right),
\end{equation}
which can be solved for the QNM frequencies by analytically continuing to the complex $\omega$ plane \cite{iyer1987black1}.
This is the generalisation of the WKB condition in Eq.~\eqref{QNM0} to the LV regime and is the main result of this work.
Indeed, if one sets \mbox{$\sqrt{r}\psi(r)\sim e^{i\int p_*(r_*) dr_*}$} in Eq.~\eqref{1d_eq}, the corresponding effective Hamiltonian is \mbox{$\mathcal{H} = \frac{1}{2}p_*^2 + \frac{1}{2}V(r_*)$}. Using this $\mathcal{H}$ in \eqref{QNM1} (and replacing $\partial_r\to\partial_{r_*}$, $\partial_p\to\partial_{p_*}$) one can then recover the original condition in Eq.~\eqref{QNM0}.
Under the assumption that the real part of the frequency is much larger than the imaginary part, condition \eqref{QNM1} yields an approximation for QNMs in terms of the light-ring frequency,
\begin{equation} \label{qnm_final}
\begin{split}
\omega^\pm_\mathrm{QNM} \simeq & \ \omega^\pm_\lr(m) - i\Lambda(m)\left(n+\frac{1}{2}\right), \\
\Lambda(m) = & \ \frac{\sqrt{-\mathrm{det}[\mathrm{d}^2\mathcal{H}]}}{|\partial_\omega\mathcal{H}|}\bigg|_\lr.
\end{split}
\end{equation}
where $\Lambda$ is the Lyapunov exponent that measures the divergence of rays moving away from the light-ring \cite{cardoso2009geodesic,torres2018waves}. 
The overtone number $n=0,1,...$ orders the QNM frequencies according to their lifetime, with the fundamental mode $n=0$ being the longest lived. 
Although this approximation was already used in \cite{torres2018waves}, we have shown here that it is consistent with a full treatment of all spatial modes in the system arising due to dispersion \footnote{This would not be the case, for example, if the out-going mode were reflected back into the vortex by a combination of a dispersion and velocity profile.}.

\section{Results} \label{sec:results}

We now use the formula in \eqref{qnm_final} to study the QNM spectrum subject to LV modifications of the form \eqref{E-M}.
To streamline the analysis, we use the fact that in our approximation, which is valid for large $\mathrm{Re}[\omega]$ and small $\mathrm{Im}[\omega]$, the real part of the overtones is the same and the imaginary part increases in increments of $\Lambda(m)$.
Furthermore, we can efficiently compare the co- and counter-rotating QNMs by plotting the latter for negative $C$ values as was done for the light-ring radius in Fig.~\ref{fig:LR}.
The QNM spectrum is therefore characterised by the plots of light-ring frequency $\omega_\lr^\pm$ and Lyapunov exponent $\Lambda(m)$ in Fig.~\ref{fig:Cdep}.
We display results for several $m$-modes, with the $\omega_\lr^\pm$ scaled by $m$ to enable an efficient comparison.
The LI results ($l^2=0$) show excellent agreement with the results of the continued fraction method (CFM) of \cite{cardoso2004qnm}, which is the gold standard method for precisely computing QNM frequencies \cite{leaver1985analytic}. 
We observe that co-rotating modes oscillate and decay faster than the counter-rotating modes, with this difference becoming more pronounced for larger rotations.
We also see that the oscillation frequency scales linearly with $m$ whilst the decay rate is $m$ independent.
This agrees with results in the literature: the oscillation frequency increases with $m$ but the lifetime shows only a weak dependence, whereas the opposite is true for variations in the overtone number $n$ \cite{leaver1985analytic,matyjasek2024accurate}. These trends emerge in the limit of large $m$ and small $n$, where the assumption of \mbox{$|\mathrm{Im}[\omega]|\ll|\mathrm{Re}[\omega]|$} is satisfied.

Switching on LV modifications, a general trend is that the co-rotating modes are affected more strongly than the counter-rotating ones.
For $l^2<0$, both the oscillation frequency and decay rate are reduced relative to the LI case, and there is a critical $C$ value where the oscillation frequency and decay rate suddenly decrease.
This $C$ value is indicated by a solid grey line and corresponds to $C^-_\star$ shown in Fig.~\ref{fig:critC}.
Hence slowly-oscillating, long-lived co-rotating modes are associated with the scattering of the $p^{+,u}$ mode pair, which happens for rapid rotations.
A physical signature of this process would be a short-wavelength mode propagating into the vortex core, compared to the usual case where the in-going mode is red-shifted.
The effect is more pronounced for higher $m$-modes, which have shorter wavelength and therefore more readily probe the scale where LV physics onsets.

For $l^2>0$, the oscillation frequency and decay rate are larger than those in the LI case, and there is a critical $C$ value (shown this time by a grey dashed line) where both suddenly increase in magnitude. This corresponds to $C^+_\star$ on Fig.~\ref{fig:critC} and marks the rotation where the light-ring radius goes to zero (see Fig.~\ref{fig:LR}).
This means that there are no co-rotating QNMs for $C>C^+_\star$ for any strength of LV modifications.
Furthermore, when LV modifications are strong (specifically when $\gamma>1/2$) there are no co-rotating QNMs at any rotation and no counter-rotating ones at small rotations in the range $|C|<|C^+_\star|$.
This implies that not all $m$-modes exhibit quasi-normal ringing for superluminal LV effects.

Although the characteristic behaviour of the QNM spectrum changes near the critical rotations shown in Fig.~\ref{fig:critC}, it is interesting to ask at what point do small deviations start to become measurable.
For the oscillation frequency, the linear corrections in $\gamma$ are given in \eqref{LR_lin}.
For the decay rate, we find numerically that Lypunov exponent is given by \mbox{$\Lambda=\Lambda_0(1+3\gamma B_\pm^2/2) + \mathcal{O}(\gamma^2)$}.
Hence, linear deviations start to become important when the product $\gamma B_\pm^2$ reaches a significant fraction of unity.
If we set this fraction to $25\%$, the critical $C$ value coincides exactly with $C^+_\star$.
Hence, corrections $\mathcal{O}(\gamma)$ contribute on the order of a few percent at a $C$ value slightly lower than this. 
In summary, if the dispersive scale is very small ($|\LV|\ll r_h$) deviations away from the LI spectrum are maximised by considering higher $m$-modes and larger ratios of $C/D$.

\section{Discussion} \label{sec:conc}

We have studied the influence of a quartic LV modification to the massless particle energy momentum relation [see Eq.~\eqref{E-M}] in the context of a specific toy model.
The model is motivated by the analogue gravity framework and consists of low frequency surface gravity waves around a draining vortex, modelling a rotating BH spacetime.
The main part of the analysis was to derive a condition to determine QNM frequencies that parallels the well-known WKB formula in \eqref{QNM0} but is equipped to handle LV modifications.
Taking into consideration extra spatial modes generated by LV effects, we arrived at the formula \eqref{QNM1} which reproduces the well-known light-ring approximation in \eqref{qnm_final} in the high frequency limit (which is the natural setting for WKB methods).
This formula may be of use in modified gravity theories such as Ho\v{r}ava-Lifshitz and Einstein-Aether gravity, which also break LI \cite{barausse2011black,barausse2013black}.

Applying this to our problem, we found that the QNM spectrum exhibits notable deviations approaching critical rotation rates that are set by the strength of the LV modifications, as well as the wave angular momentum $m$.
We found that sub- and superluminal modifications led to respective decreases and increases in both the oscillation frequency and decay rates of QNMs.
In the subluminal case, this was most noticeable at high rotations for co-rotating modes, where one of the LV modes forms the in-going part of the QNM.
In the superluminal case, we found that there are no co-rotating QNMs above a maximum rotation and that, for strong enough LV modifications, counter-rotating QNMs were also absent at low rotation.
A general trend is that deviations from the LI spectrum can be significant for co-rotating modes when LV effects are weak provided rotations are high, whereas modifications to the counter-rotating spectrum require strong LV effects.

QNM oscillations have already been measured in the strongly dispersive regime in classical fluid experiments \cite{torres2017rotational}.
Recent experiments in superfluid $^4$He showed promising ringdown signatures \cite{svanvcara2024rotating} and resonant peaks characteristic of QNMs in finite size set-ups (see \cite{solidoro2024quasinormal}) were detected in \cite{smaniotto2025black}.
Other recent experiments operate in the regime where the rotating curved spacetime geometry is generated by a quantum fluid with discretised angular momentum \cite{braidotti2022measurement,delhom2024entanglement} and QNMs have also been shown to arise in optical solitons \cite{burgess2024quasinormal}, where optical control can enable the measurement of quantum correlations.
Furthermore, the decay of quantum vortices in cold atomic gases \cite{shin2004dynamical} is related to superradiance \cite{giacomelli2020ergoregion,patrick2022quantum,patrick2024quantum}, which can be interpreted within the effective spacetime picture.
Altogether, considerations of such analogue models could be used to address how quantum effects in the background (and therefore the effective BH spacetime metric) imprint on wave scattering processes such as QNM ringing.
Since quantum effects in these systems typically break the effective Lorentz symmetry of low energy quasiparticles, LV effects will undoubtedly be an important part of the story.
This is an interesting future direction worth exploring.

Our conclusions may also have theoretical implications.
It has been suggested that higher QNM overtones may be related to a BH's ability to absorb quantised energy packets, which has been argued to yield a quantisation rule for the BH's area \cite{hod1998bohr,maggiore2008physical}.
If Lorentz symmetry is not preserved at high energies, then higher overtones will be sensitive and their frequencies affected.
This could be investigated using the monodromy technique discussed in \cite{berti2009review} which yields an asymptotic formula for QNM frequencies in the large $n$ limit.

In terms of real black holes, this would be a minuscule effect.
If we take $|\LV|=l_P$ to be the Planck length, then for a solar mass BH ($M_\odot\sim 10^{30}\mathrm{kg}$) the coefficient of the LV modification would be $l^2_P/r_h^2 \sim 10^{-76}$.
Even for the smallest primordial BHs which would not have evaporated away by today \cite{carr2021constraints} ($M_\mathrm{PBH}\sim 10^{12}\mathrm{kg}$) this number is still tiny $l_P^2/r_h^2\sim 10^{-36}$.
However, we have seen that LV effects on the QNM spectrum can be amplified by higher rotations and higher angular momentum numbers (in our case $m$).
Kerr BHs are different to the model we studied since (a) they live in three spatial dimension, (b) their horizon radius decreases with rotation and (c) they have a maximum angular momentum at fixed mass given by the extremal limit.
It would be interesting to investigate how these effects modify our conclusions to see whether LV effects can influence QNMs within a detectable range.
Even if not detectable astrophysically, the LV effects we described become most relevant for Planckian BHs with $r_h \sim \mathcal{O}(l_p)$ and, as such, may have consequences for the end point of BH evaporation.

\textbf{Acknowledgements}. The authors thank Patrik \v{S}van\v{c}ara, Pietro Smaniotto, Ruth Gregory, and Silke Weinfurtner for stimulating discussions.
S.~P. extends appreciation to the Science and Technology Facilities Council for their generous support in Quantum Simulators for Fundamental Physics (QSimFP), (ST/T005858/1), and the Leverhulme Trust (Grant No. RPG-2016-233) for support during an earlier version of the project.
L.~S. gratefully acknowledges the support of the
Leverhulme Research Leadership Award (RL-2019-020).

\appendix

\section{Validity of $\LVT<0$} \label{app:restrict}

Here we discuss the range of applicability of \eqref{disp1} when using $\LVT<0$. The curve $\omega_D^+$ ($\omega_D^-$) defined in Eq.~\eqref{disp_UL} displays a maximum (minimum) in the $p<0$ ($p>0$) region in the $r\to\infty$ limit, which means that above (below) a critical frequency there will be no asymptotically propagating modes.
This comes from the fact that the right-hand side of \eqref{disp1} has two maxima in $k$, whereas the true dispersion relation for capillary-gravity waves does not have this feature, see e.g. \cite{patrick2024primer}.
To regulate this spurious behaviour, let $r=L$ be a large radius which we define as the asymptotic region (that is, there should be no further scattering for $r>L$).
Then, provided we restrict ourselves to frequencies satisfying,
\begin{equation} \label{restrict}
\mathrm{min}[\omega_D^-(L,p<0)]<\omega<\mathrm{max}[\omega_D^+(L,p>0)],
\end{equation}
when $\LVT<0$, we can avoid the spurious effects resulting from the truncation of Eq.~\eqref{disp1} to quartic order.

\section{Scattering around a saddle point} \label{app:saddle}

We expand the effective Hamiltonian $\mathcal{H}$ \eqref{effH2} near the saddle point defined by \eqref{sp_def}. 
Under the assumption that \mbox{$R=r-r_\sp$} and \mbox{$P=p-p_\sp$} are both small, we expand to quadratic order,
\begin{equation}
\mathcal{H} = H + \frac{H_{pp}}{2}P^2 + H_{rp}RP + \frac{H_{rr}}{2}R^2
\end{equation}
where for notational convenience we have introduced $H=\mathcal{H}_\sp$, $H_p = \partial_p\mathcal{H}|_\sp$ and similarly for $r$ derivatives.
When we complete the square, we obtain
\begin{equation}
\begin{split}
\mathcal{H} = H + \frac{H_{pp}}{2}\left(P+\frac{H_{pr}}{H_{pp}}R\right)^2 +\frac{\Hess}{2H_{pp}}R^2 = 0
\end{split}
\end{equation}
where we have defined the determinant of the Hessian matrix,
\begin{equation}
\Hess = H_{pp}H_{rr}-H_{pr}^2.
\end{equation}
Now, we want to promote $\mathcal{H}$ to an operator to find the local expansion of the wave equation near the saddle point.
First, consider defining a new field $\psi$ related to the velocity potential perturbation $\phi$ by,
\begin{equation}
\begin{split}
\phi = & \ e^{im\theta-i\omega t}\exp\left(i\int^R_{0}f(R')dR'\right) \psi(r), \\
f(R) = & \ p_\sp - \frac{H_{pr}}{H_{pp}}R.
\end{split}
\end{equation}
In shallow water, this is the same redefinition of the field as in \eqref{expans2} but with a linear expansion of the integral in the phase term and the choice that this phase goes to zero at $r_\sp$.

The WKB approximation on $\psi$ (consistent with \mbox{$\phi\sim e^{i\int p dr}$}) is \mbox{$\psi\sim e^{i\int P' dR}$}.
This gives,
\begin{equation}
\mathcal{H} = H + \frac{H_{pp}}{2}{P'}^2 + \frac{\Hess}{2H_{pp}}R^2.
\end{equation}
Defining,
\begin{equation} \label{bXK_def}
\begin{split}
b = & \ \frac{\epsilon H}{\sqrt{-\Hess}}, \\
K = & \ \frac{(\epsilon H_{pp})^\frac{1}{2}P'}{(-\Hess)^\frac{1}{4}}, \qquad X = \frac{(-\Hess)^\frac{1}{4}R}{(\epsilon H_{pp})^\frac{1}{2}},
\end{split}
\end{equation}
where the factor $\epsilon = 1$ ($\epsilon=-1$) for case 1 (case 2) in Fig.~\ref{fig:saddle}, one finds,
\begin{equation} \label{H_KX}
\begin{split}
\mathcal{H} = \epsilon\sqrt{-\Hess}\left[\frac{K^2}{2}-\frac{X^2}{2} +b\right].
\end{split}
\end{equation}
The WKB solutions are given by,
\begin{equation}
\psi \propto \frac{a}{|K|^\frac{1}{2}}e^{i\int K dX}
\end{equation}
where $a$ is a $\mathbb{C}$-number and the proportionality factor is $|f(-\Hess)^{-1/4}|^{1/2}$ which is the same for both modes.
The prefactor $|K|^{-1/2}$ is obtained by solving the next to leading order equation for the amplitude in the WKB expansion, see e.g. \cite{torres2018waves,patrick2020superradiance}.
We want to know the functional form of these modes a reasonable distance away from the saddle point on either side. For this discussion, we take $\omega$ to be real. Later we will generalise to a small imaginary part.
$\mathcal{H}=0$ can be solved for the local values of the radial wavevector,
\begin{equation}
K = \pm \sqrt{X^2-2b}.
\end{equation}
When $b>0$ there are two real turning points at \mbox{$X_\pm=\pm\sqrt{2b}$} whilst for $b<0$ there are no real turning points and $K$ is real everywhere in the vicinity of the saddle point. 
For $b>0$, we search for the form of the modes outside the turning point $|X|\gg|X_\pm|$ where we expect the WKB approximation to be valid.
To this end, we evaluate the asymptotic limit of the phase integral,
\begin{equation}
\int_{X_\pm}^{X}dX'\sqrt{{X'}^2-2b} \sim \pm\left[\frac{X^2}{2} + \frac{b}{2}\log\left(\frac{b}{2eX^2}\right)\right],
\end{equation}
where the $\pm$ is given by $\mathrm{sgn}(X)$ and we have neglected terms $\mathcal{O}(X^{-2})$.
For $b<0$, the lower limit of the integral goes to zero and we find,
\begin{equation} \label{WKBint}
\int_0^{X}dX'\sqrt{{X'}^2-2b} \sim \pm\left[\frac{X^2}{2} + \frac{b}{2}\log\left(\frac{-b}{2eX^2}\right)\right].
\end{equation}
To determine which mode is ingoing or outgoing, we compute the group velocity $v_g^r=\Omega^{-1}\partial_p\mathcal{H}$, whose sign is $\mathrm{sgn}(\epsilon \Omega K)$. In both cases, the factor in the amplitude is $|K|^{-1/2}=|X|^{-1/2}$.
Collectively, we can evaluate the WKB modes at the locations,
\begin{equation}
X_> = \begin{cases}
X_+ \\
0^+
\end{cases} \quad
X_< = \begin{cases}
X_- \quad b>0 \\
0^- \quad \ b<0
\end{cases},
\end{equation}
where the $0^\pm$ indicates that the limit zero is approached from above or below.
Thus, the WKB solution in the two regions may be written,
\begin{equation} \label{WKBsad}
\begin{split}
\psi(X>0) \propto a^+_>\psi^+_> + a^-_>\psi^-_>, \\
\psi(X<0) \propto a^+_<\psi^+_< + a^-_<\psi^-_<, \\
\end{split}
\end{equation}
where,
\begin{equation} \label{WKBmodes}
\begin{split}
\psi^+_> = & \ \psi^-_< = |\sqrt{2}X|^{-ib-\frac{1}{2}}e^{iX^2/2}|b|^\frac{ib}{2}e^{-\frac{ib}{2}}, \\
\psi^-_> = & \ \psi^+_< = |\sqrt{2}X|^{ib-\frac{1}{2}}e^{-iX^2/2}|b|^{-\frac{ib}{2}}e^\frac{ib}{2}.
\end{split}
\end{equation}
Note the term involving $|b|$ assumes $b$ to be real since the distinction between having/not having real turning points assumes $\omega\in\mathbb{R}$. Later we address how to extend to complex $\omega$.

\begin{figure*} 
\centering
\includegraphics[width=\linewidth]{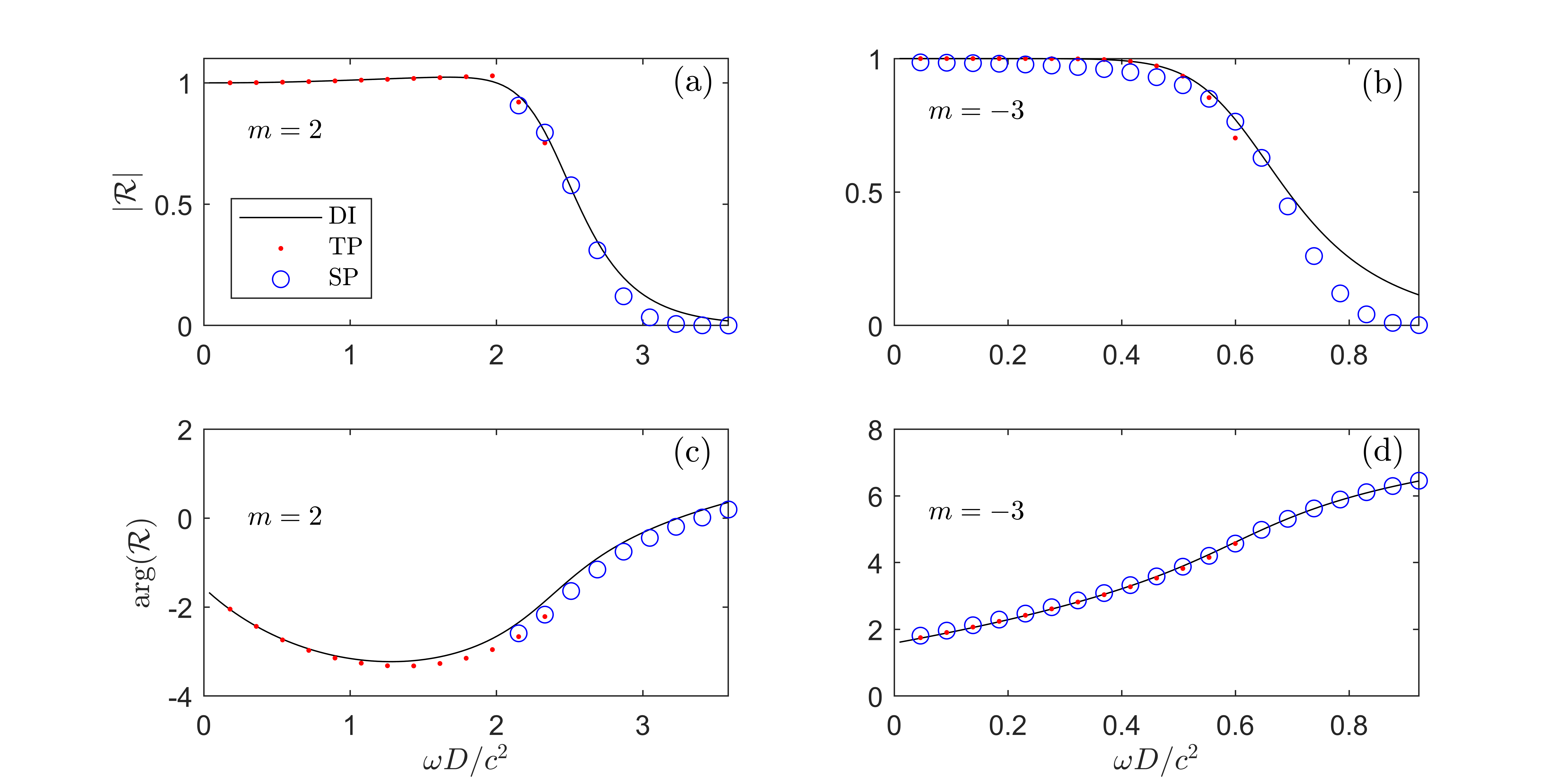}
\caption{The shallow water reflection coefficient $\mathcal{R}$ at infinity of an unbounded draining vortex. The magnitude and phase of $\mathcal{R}$ are shown in panels (a,b) and (c,d) respectively for $m=2,-3$. The black line is the exact numerical value obtained through direct integration (DI) whereas the red points and blue circles correspond to the WKB estimates using the turning point (TP) and saddle point (SP) methods respectively. Both methods show good agreement, although only the SP method can be applied above the light-ring frequency $\omega_\lr$, since the TP method requires real turning points.
} \label{fig:shal_refl}
\end{figure*}

The goal now is to match the solution \eqref{WKBsad} onto an asymptotic expansion of the exact solution in the vicinity of the turning point and in doing so find a transfer matrix connecting the amplitudes $a^\pm_<$ with $a^\pm_>$.
Promoting $K\to -i\partial_X$ in \eqref{H_KX}, we obtain the local form of the wave equation,
\begin{equation}
	\left(\frac{1}{2}\partial_X^2 + \frac{X^2}{2}-b\right)\psi = 0.
\end{equation}
The solution of this equation can be written as a sum of parabolic cylinder functions \cite{torres2020estimate},
\begin{equation} \label{exact_PCF}
\psi = C_1U(ib,\sqrt{2}Xe^{-i\pi/4}) + C_2U(-ib,\sqrt{2}Xe^{i\pi/4}),
\end{equation}
where $C_{1,2}$ are two integration constants.
The asymptotic expansion of these functions can be written as,
\begin{widetext}
\begin{equation} \label{U_asymp}
\begin{split}
U(a,z) \sim & \ e^{-z^2/4}z^{-a-1/2}, \qquad |\mathrm{arg} z|<3\pi/4, \\
\sim & \ e^{-z^2/4}z^{-a-1/2} \pm \frac{\sqrt{2\pi}i}{\Gamma(\frac{1}{2}+a)}e^{\mp i\pi a} e^{z^2/4}z^{a-1/2}, \qquad \pi/4 < \pm\mathrm{arg} z < 5\pi/4.
\end{split}
\end{equation}
\end{widetext}
Taking $X$ to be real, we can evaluate the phase of function arguments in \eqref{exact_PCF} to decide which expansion to apply \footnote{Note that for $\omega\in\mathbb{C}$, the prefactors in \eqref{bXK_def} make $X$ complex. However, if the imaginary part is small then the phase of the arguments only receive a small correction. In that case, the asymptotic expansions derived for $\omega\in\mathbb{R}$ (which apply in a particular region of the complex plane) will still be valid, since a perturbative $\mathrm{Im}[\omega]$ will not displace the function out of this region.}. For $X>0$, the first line of \eqref{U_asymp} is relevant while the second applies to $X<0$.
Using the asymptotic expansions each side of the saddle point, we find,
\begin{equation} \label{Usad}
\begin{split}
\psi(X>0) = & \ C_1U_{1>} + C_2U_{2>}, \\
\psi(X<0) = & \ C_1U_{1<} + C_2U_{2<},
\end{split}
\end{equation}
where we have written the asymptotic expansions in terms of the WKB modes in \eqref{WKBmodes},
\begin{equation}
\begin{split}
U_{1>} = & \ \tau^\frac{1}{4}\alpha\psi^+_>, \\
U_{2>} = & \ \tau^\frac{1}{4}\alpha^{-1}\psi^-_>, \\
U_{1<} = & \ \tau^{-\frac{3}{4}}\left(\beta\alpha^{-1}\psi^+_< - i\alpha\psi^-_< \right), \\
U_{2<} = & \ \tau^{-\frac{3}{4}}\left(i\alpha^{-1}\psi^+_< +\beta^*\alpha\psi^-_< \right), \\
\end{split}
\end{equation}
and we have also defined,
\begin{equation} \label{sad_params}
\tau = e^{-\pi b}, \quad \beta = \frac{\sqrt{2\pi i\tau}}{\Gamma(\frac{1}{2}+ib)}, \quad \alpha = |b|^{-\frac{ib}{2}}e^{\frac{ib}{2}+\frac{i\pi}{8}}.
\end{equation}
Comparing \eqref{Usad} with \eqref{WKBsad}, one finds the following connection matrix relating WKB amplitudes either side of the saddle point,
\begin{equation} \label{N_sp}
\begin{pmatrix}
a^+_< \\ a^-_<
\end{pmatrix} = \tau^{-1}\begin{pmatrix}
\beta \alpha^{-2} & i \\ -i & \beta^*\alpha^2
\end{pmatrix} \begin{pmatrix}
a^+_>\\ a^-_>
\end{pmatrix}.
\end{equation}
The relation between the amplitudes of $\phi$ in \eqref{RL_transfer} contains an additional phase due to the factor $e^{i\int f(R) dR}$ when converting between $\psi$ and $\phi$.

Let us test this approximation in the shallow water regime. The solutions to \eqref{1d_eq} which are incoming from infinity and in-going on $r=r_h$ have the form,
\begin{equation} \label{psi_asymp}
\sqrt{r}\psi \sim \begin{cases}
e^{-i\omega r_*} + \mathcal{R}e^{i\omega r_*}, \quad & r_*\to\infty \\
\mathcal{T} e^{-i\tilde{\omega}_h r_*}, & r_*\to-\infty
\end{cases}
\end{equation}
where $\tilde{\omega}_h=\tilde{\omega}(r_h)$ and $\mathcal{R}$ and $\mathcal{T}$ are the reflection and transmission coefficients.
Let us calculate $\mathcal{R}$ within the WKB approximation.
The relation between the constant part of WKB amplitudes of the $\psi$ is,
\begin{equation} \label{A_matrix}
\begin{split}
\begin{pmatrix}
e^{iS_{c<}}a^+_c \\ e^{-iS_{c<}}a^-_c
\end{pmatrix} = & \ \mathcal{N} \begin{pmatrix}
e^{-iS_{>L}}a^+_L \\ e^{iS_{>L}}a^-_L
\end{pmatrix}, \\
S_{ab} = & \ \int^{r_b}_{r_a} \frac{c\sqrt{-V} dr}{c^2-v_r^2},
\end{split}
\end{equation}
where $r_c$ ($L$) is a location deep in the vortex core (far outside the vortex) such that no further scattering occurs for $r<r_c$ ($r>L$) and $\mathcal{N}$ is a $2\times 2$ transfer matrix relating the amplitudes of the modes at $x_<$ (on the left) to those at $x_>$ (on the right).
The form of this depends on the frequency and the approximation we are using.
For $\tilde{\omega}_h/\omega>0$, the mode labelled $p^+$ ($p^-$) is out-going (in-going) at $r_c$ and we have,
\begin{equation}
\mathcal{N} = \begin{pmatrix}
1/\mathbf{T} & -\mathbf{R}/\mathbf{T}, \\
-\mathbf{R}^*/\mathbf{T}^* & 1/\mathbf{T}^*
\end{pmatrix},
\end{equation}
where $\mathbf{R}$ ($\mathbf{T}$) is the local reflection (transmission) coefficient defined at $x_>$ ($x_<$).
These can be calculated using the saddle point connection matrix in \eqref{N_sp} or using the turning point connection matrix (see e.g. Eq.~(52) of \cite{patrick2021rotational}),
\begin{equation}
\begin{split}
\mathbf{R}_\sp = -i\frac{\alpha^2}{\beta}, & \ \quad \mathbf{R}_\tp = -i\frac{1-e^{-2\mathbf{S}}/4}{1+e^{-2\mathbf{S}}/4}, \\
\mathbf{T}_\sp = \tau\frac{\alpha^2}{\beta}, & \ \quad \mathbf{T}_\tp = \frac{e^{-\mathbf{S}}}{1+e^{-2\mathbf{S}}/4}.
\end{split}
\end{equation}
where $\mathbf{S} = |\mathrm{Im}[S_{<>}]|$.
The turning point expressions for $\mathbf{R}$ and $\mathbf{T}$ require real turning points and work best when they are well separated, whereas the saddle point results can deal with complex turning points but work best when they are close together.
For $\tilde{\omega}_h/\omega<0$, the mode labelled $p^+$ ($p^-$) is in-going (out-going) at $r_c$ and we have,
\begin{equation}
\mathcal{N} = \begin{pmatrix}
\mathbf{R}^*/\mathbf{T}^* & -1/\mathbf{T}^* \\
-1/\mathbf{T} & \mathbf{R}/\mathbf{T}
\end{pmatrix}.
\end{equation}
In this case, the turning points are well-separated and we have,
\begin{equation}
\begin{split}
\mathbf{R}_\tp = & \ -i\frac{1+e^{-2\mathbf{S}}/4}{1-e^{-2\mathbf{S}}/4} \\
\mathbf{T}_\tp = & \ -i\frac{e^{-\mathbf{S}}}{1-e^{-2\mathbf{S}}/4}
\end{split}
\end{equation}
Comparing with \eqref{psi_asymp}, we obtain the reflection coefficient at infinity (i.e. the one measured far away from the vortex core assuming the in- and out-going waves are plane waves),
\begin{equation} \label{RT_asymp}
\mathcal{R} = \mathbf{R}\underset{L\to\infty}{\mathrm{lim}}e^{2i(S_{<L}-\omega L/c)},
\end{equation}
Note that the magnitude of these is just given by the local coefficients, whilst the term under the $L\to\infty$ limit tends to a constant phase contribution.
The results of \eqref{RT_asymp} using both the saddle point and turning point methods are illustrated on Fig.~\ref{fig:shal_refl} and compared with an exact numerical result obtained from a direct integration method (described in e.g. \cite{churilov2018scattering,patrick2020analogy}).
All methods show good agreement in both the magnitude and phase of $\mathcal{R}$.

\section*{References}
\bibliography{final.bbl}

\begin{thebibliography}{65}%
\makeatletter
\providecommand \@ifxundefined [1]{%
 \@ifx{#1\undefined}
}%
\providecommand \@ifnum [1]{%
 \ifnum #1\expandafter \@firstoftwo
 \else \expandafter \@secondoftwo
 \fi
}%
\providecommand \@ifx [1]{%
 \ifx #1\expandafter \@firstoftwo
 \else \expandafter \@secondoftwo
 \fi
}%
\providecommand \natexlab [1]{#1}%
\providecommand \enquote  [1]{``#1''}%
\providecommand \bibnamefont  [1]{#1}%
\providecommand \bibfnamefont [1]{#1}%
\providecommand \citenamefont [1]{#1}%
\providecommand \href@noop [0]{\@secondoftwo}%
\providecommand \href [0]{\begingroup \@sanitize@url \@href}%
\providecommand \@href[1]{\@@startlink{#1}\@@href}%
\providecommand \@@href[1]{\endgroup#1\@@endlink}%
\providecommand \@sanitize@url [0]{\catcode `\\12\catcode `\$12\catcode `\&12\catcode `\#12\catcode `\^12\catcode `\_12\catcode `\%12\relax}%
\providecommand \@@startlink[1]{}%
\providecommand \@@endlink[0]{}%
\providecommand \url  [0]{\begingroup\@sanitize@url \@url }%
\providecommand \@url [1]{\endgroup\@href {#1}{\urlprefix }}%
\providecommand \urlprefix  [0]{URL }%
\providecommand \Eprint [0]{\href }%
\providecommand \doibase [0]{https://doi.org/}%
\providecommand \selectlanguage [0]{\@gobble}%
\providecommand \bibinfo  [0]{\@secondoftwo}%
\providecommand \bibfield  [0]{\@secondoftwo}%
\providecommand \translation [1]{[#1]}%
\providecommand \BibitemOpen [0]{}%
\providecommand \bibitemStop [0]{}%
\providecommand \bibitemNoStop [0]{.\EOS\space}%
\providecommand \EOS [0]{\spacefactor3000\relax}%
\providecommand \BibitemShut  [1]{\csname bibitem#1\endcsname}%
\let\auto@bib@innerbib\@empty
\bibitem [{\citenamefont {Berti}\ \emph {et~al.}(2009)\citenamefont {Berti}, \citenamefont {Cardoso},\ and\ \citenamefont {Starinets}}]{berti2009review}%
  \BibitemOpen
  \bibfield  {author} {\bibinfo {author} {\bibfnamefont {E.}~\bibnamefont {Berti}}, \bibinfo {author} {\bibfnamefont {V.}~\bibnamefont {Cardoso}},\ and\ \bibinfo {author} {\bibfnamefont {A.~O.}\ \bibnamefont {Starinets}},\ }\bibfield  {title} {\bibinfo {title} {Quasinormal modes of black holes and black branes},\ }\href@noop {} {\bibfield  {journal} {\bibinfo  {journal} {Classical and Quantum Gravity}\ }\textbf {\bibinfo {volume} {26}},\ \bibinfo {pages} {163001} (\bibinfo {year} {2009})}\BibitemShut {NoStop}%
\bibitem [{\citenamefont {Gamow}(1928)}]{gamow1928quantentheorie}%
  \BibitemOpen
  \bibfield  {author} {\bibinfo {author} {\bibfnamefont {G.}~\bibnamefont {Gamow}},\ }\bibfield  {title} {\bibinfo {title} {Zur quantentheorie des atomkernes},\ }\href@noop {} {\bibfield  {journal} {\bibinfo  {journal} {Zeitschrift f{\"u}r Physik}\ }\textbf {\bibinfo {volume} {51}},\ \bibinfo {pages} {204} (\bibinfo {year} {1928})}\BibitemShut {NoStop}%
\bibitem [{\citenamefont {Leung}\ \emph {et~al.}(1994)\citenamefont {Leung}, \citenamefont {Liu},\ and\ \citenamefont {Young}}]{leung1994completeness}%
  \BibitemOpen
  \bibfield  {author} {\bibinfo {author} {\bibfnamefont {P.~T.}\ \bibnamefont {Leung}}, \bibinfo {author} {\bibfnamefont {S.~Y.}\ \bibnamefont {Liu}},\ and\ \bibinfo {author} {\bibfnamefont {K.}~\bibnamefont {Young}},\ }\bibfield  {title} {\bibinfo {title} {Completeness and orthogonality of quasinormal modes in leaky optical cavities},\ }\href@noop {} {\bibfield  {journal} {\bibinfo  {journal} {Phys. Rev. A}\ }\textbf {\bibinfo {volume} {49}},\ \bibinfo {pages} {3057} (\bibinfo {year} {1994})}\BibitemShut {NoStop}%
\bibitem [{\citenamefont {Kristensen}\ \emph {et~al.}(2015)\citenamefont {Kristensen}, \citenamefont {Ge},\ and\ \citenamefont {Hughes}}]{kristensen2015normalization}%
  \BibitemOpen
  \bibfield  {author} {\bibinfo {author} {\bibfnamefont {P.~T.}\ \bibnamefont {Kristensen}}, \bibinfo {author} {\bibfnamefont {R.}~\bibnamefont {Ge}},\ and\ \bibinfo {author} {\bibfnamefont {S.}~\bibnamefont {Hughes}},\ }\bibfield  {title} {\bibinfo {title} {Normalization of quasinormal modes in leaky optical cavities and plasmonic resonators},\ }\href@noop {} {\bibfield  {journal} {\bibinfo  {journal} {Phys. Rev. A}\ }\textbf {\bibinfo {volume} {92}},\ \bibinfo {pages} {053810} (\bibinfo {year} {2015})}\BibitemShut {NoStop}%
\bibitem [{\citenamefont {Abbott}\ \emph {et~al.}(2016)\citenamefont {Abbott}, \citenamefont {Abbott}, \citenamefont {Abbott}, \citenamefont {Abernathy}, \citenamefont {Acernese}, \citenamefont {Ackley}, \citenamefont {Adams}, \citenamefont {Adams}, \citenamefont {Addesso}, \citenamefont {Adhikari} \emph {et~al.}}]{abbott2016observation}%
  \BibitemOpen
  \bibfield  {author} {\bibinfo {author} {\bibfnamefont {B.~P.}\ \bibnamefont {Abbott}}, \bibinfo {author} {\bibfnamefont {R.}~\bibnamefont {Abbott}}, \bibinfo {author} {\bibfnamefont {T.~D.}\ \bibnamefont {Abbott}}, \bibinfo {author} {\bibfnamefont {M.~R.}\ \bibnamefont {Abernathy}}, \bibinfo {author} {\bibfnamefont {F.}~\bibnamefont {Acernese}}, \bibinfo {author} {\bibfnamefont {K.}~\bibnamefont {Ackley}}, \bibinfo {author} {\bibfnamefont {C.}~\bibnamefont {Adams}}, \bibinfo {author} {\bibfnamefont {T.}~\bibnamefont {Adams}}, \bibinfo {author} {\bibfnamefont {P.}~\bibnamefont {Addesso}}, \bibinfo {author} {\bibfnamefont {R.~X.}\ \bibnamefont {Adhikari}}, \emph {et~al.},\ }\bibfield  {title} {\bibinfo {title} {Observation of gravitational waves from a binary black hole merger},\ }\href@noop {} {\bibfield  {journal} {\bibinfo  {journal} {Phys. Rev. Lett.}\ }\textbf {\bibinfo {volume} {116}},\ \bibinfo {pages} {061102} (\bibinfo {year} {2016})}\BibitemShut {NoStop}%
\bibitem [{\citenamefont {Kokkotas}\ and\ \citenamefont {Schmidt}(1999)}]{kokkotas1999stars}%
  \BibitemOpen
  \bibfield  {author} {\bibinfo {author} {\bibfnamefont {K.~D.}\ \bibnamefont {Kokkotas}}\ and\ \bibinfo {author} {\bibfnamefont {B.~G.}\ \bibnamefont {Schmidt}},\ }\bibfield  {title} {\bibinfo {title} {{Quasinormal modes of stars and black holes}},\ }\href {https://doi.org/10.12942/lrr-1999-2} {\bibfield  {journal} {\bibinfo  {journal} {Living Reviews in Relativity}\ }\textbf {\bibinfo {volume} {2}},\ \bibinfo {pages} {2} (\bibinfo {year} {1999})},\ \Eprint {https://arxiv.org/abs/gr-qc/9909058} {arXiv:gr-qc/9909058 [gr-qc]} \BibitemShut {NoStop}%
\bibitem [{\citenamefont {Dreyer}\ \emph {et~al.}(2004)\citenamefont {Dreyer}, \citenamefont {Kelly}, \citenamefont {Krishnan}, \citenamefont {Garrison},\ and\ \citenamefont {Lopez-Aleman}}]{dreyer2004black}%
  \BibitemOpen
  \bibfield  {author} {\bibinfo {author} {\bibfnamefont {O.}~\bibnamefont {Dreyer}}, \bibinfo {author} {\bibfnamefont {B.}~\bibnamefont {Kelly}}, \bibinfo {author} {\bibfnamefont {L.~S.}\ \bibnamefont {Krishnan}, \bibfnamefont {B.and~Finn}}, \bibinfo {author} {\bibfnamefont {D.}~\bibnamefont {Garrison}},\ and\ \bibinfo {author} {\bibfnamefont {R.}~\bibnamefont {Lopez-Aleman}},\ }\bibfield  {title} {\bibinfo {title} {Black-hole spectroscopy: testing general relativity through gravitational-wave observations},\ }\href@noop {} {\bibfield  {journal} {\bibinfo  {journal} {Class. Quant. Grav.}\ }\textbf {\bibinfo {volume} {21}},\ \bibinfo {pages} {787} (\bibinfo {year} {2004})}\BibitemShut {NoStop}%
\bibitem [{\citenamefont {Leung}\ \emph {et~al.}(1997)\citenamefont {Leung}, \citenamefont {Liu}, \citenamefont {Suen}, \citenamefont {Tam},\ and\ \citenamefont {Young}}]{leung1997quasinormal}%
  \BibitemOpen
  \bibfield  {author} {\bibinfo {author} {\bibfnamefont {P.~T.}\ \bibnamefont {Leung}}, \bibinfo {author} {\bibfnamefont {Y.~T.}\ \bibnamefont {Liu}}, \bibinfo {author} {\bibfnamefont {W.~M.}\ \bibnamefont {Suen}}, \bibinfo {author} {\bibfnamefont {C.~Y.}\ \bibnamefont {Tam}},\ and\ \bibinfo {author} {\bibfnamefont {K.}~\bibnamefont {Young}},\ }\bibfield  {title} {\bibinfo {title} {Quasinormal modes of dirty black holes},\ }\href@noop {} {\bibfield  {journal} {\bibinfo  {journal} {Phys. Rev. Lett.}\ }\textbf {\bibinfo {volume} {78}},\ \bibinfo {pages} {2894} (\bibinfo {year} {1997})}\BibitemShut {NoStop}%
\bibitem [{\citenamefont {Barausse}\ \emph {et~al.}(2014)\citenamefont {Barausse}, \citenamefont {Cardoso},\ and\ \citenamefont {Pani}}]{barausse2014can}%
  \BibitemOpen
  \bibfield  {author} {\bibinfo {author} {\bibfnamefont {E.}~\bibnamefont {Barausse}}, \bibinfo {author} {\bibfnamefont {V.}~\bibnamefont {Cardoso}},\ and\ \bibinfo {author} {\bibfnamefont {P.}~\bibnamefont {Pani}},\ }\bibfield  {title} {\bibinfo {title} {Can environmental effects spoil precision gravitational-wave astrophysics?},\ }\href@noop {} {\bibfield  {journal} {\bibinfo  {journal} {Phys. Rev. D}\ }\textbf {\bibinfo {volume} {89}},\ \bibinfo {pages} {104059} (\bibinfo {year} {2014})}\BibitemShut {NoStop}%
\bibitem [{\citenamefont {Cannizzaro}\ \emph {et~al.}(2024)\citenamefont {Cannizzaro}, \citenamefont {Spieksma}, \citenamefont {Cardoso},\ and\ \citenamefont {Ikeda}}]{cannizzaro2024impact}%
  \BibitemOpen
  \bibfield  {author} {\bibinfo {author} {\bibfnamefont {E.}~\bibnamefont {Cannizzaro}}, \bibinfo {author} {\bibfnamefont {T.~F.~M.}\ \bibnamefont {Spieksma}}, \bibinfo {author} {\bibfnamefont {V.}~\bibnamefont {Cardoso}},\ and\ \bibinfo {author} {\bibfnamefont {T.}~\bibnamefont {Ikeda}},\ }\bibfield  {title} {\bibinfo {title} {Impact of a plasma on the relaxation of black holes},\ }\href@noop {} {\bibfield  {journal} {\bibinfo  {journal} {Phys. Rev. D}\ }\textbf {\bibinfo {volume} {110}},\ \bibinfo {pages} {L021302} (\bibinfo {year} {2024})}\BibitemShut {NoStop}%
\bibitem [{\citenamefont {Jaramillo}\ \emph {et~al.}(2021)\citenamefont {Jaramillo}, \citenamefont {Macedo},\ and\ \citenamefont {Sheikh}}]{jaramillo2021pseudospectrum}%
  \BibitemOpen
  \bibfield  {author} {\bibinfo {author} {\bibfnamefont {J.~L.}\ \bibnamefont {Jaramillo}}, \bibinfo {author} {\bibfnamefont {R.~P.}\ \bibnamefont {Macedo}},\ and\ \bibinfo {author} {\bibfnamefont {L.~A.}\ \bibnamefont {Sheikh}},\ }\bibfield  {title} {\bibinfo {title} {Pseudospectrum and black hole quasinormal mode instability},\ }\href@noop {} {\bibfield  {journal} {\bibinfo  {journal} {Phys. Rev. X}\ }\textbf {\bibinfo {volume} {11}},\ \bibinfo {pages} {031003} (\bibinfo {year} {2021})}\BibitemShut {NoStop}%
\bibitem [{\citenamefont {Cheung}\ \emph {et~al.}(2022)\citenamefont {Cheung}, \citenamefont {Destounis}, \citenamefont {Macedo}, \citenamefont {Berti},\ and\ \citenamefont {Cardoso}}]{cheung2022destabilizing}%
  \BibitemOpen
  \bibfield  {author} {\bibinfo {author} {\bibfnamefont {M.~H.}\ \bibnamefont {Cheung}}, \bibinfo {author} {\bibfnamefont {K.}~\bibnamefont {Destounis}}, \bibinfo {author} {\bibfnamefont {R.~P.}\ \bibnamefont {Macedo}}, \bibinfo {author} {\bibfnamefont {E.}~\bibnamefont {Berti}},\ and\ \bibinfo {author} {\bibfnamefont {V.}~\bibnamefont {Cardoso}},\ }\bibfield  {title} {\bibinfo {title} {Destabilizing the fundamental mode of black holes: the elephant and the flea},\ }\href@noop {} {\bibfield  {journal} {\bibinfo  {journal} {Phys. Rev. Lett.}\ }\textbf {\bibinfo {volume} {128}},\ \bibinfo {pages} {111103} (\bibinfo {year} {2022})}\BibitemShut {NoStop}%
\bibitem [{\citenamefont {Mattingly}(2005)}]{mattingly2005modern}%
  \BibitemOpen
  \bibfield  {author} {\bibinfo {author} {\bibfnamefont {D.}~\bibnamefont {Mattingly}},\ }\bibfield  {title} {\bibinfo {title} {Modern tests of lorentz invariance},\ }\href@noop {} {\bibfield  {journal} {\bibinfo  {journal} {Liv. Rev. Rel.}\ }\textbf {\bibinfo {volume} {8}},\ \bibinfo {pages} {1} (\bibinfo {year} {2005})}\BibitemShut {NoStop}%
\bibitem [{\citenamefont {Amelino-Camelia}(2002)}]{amelino2002relativity}%
  \BibitemOpen
  \bibfield  {author} {\bibinfo {author} {\bibfnamefont {G.}~\bibnamefont {Amelino-Camelia}},\ }\bibfield  {title} {\bibinfo {title} {Relativity in spacetimes with short-distance structure governed by an observer-independent {(Planckian)} length scale},\ }\href@noop {} {\bibfield  {journal} {\bibinfo  {journal} {Int. J. Mod. Phys. D}\ }\textbf {\bibinfo {volume} {11}},\ \bibinfo {pages} {35} (\bibinfo {year} {2002})}\BibitemShut {NoStop}%
\bibitem [{\citenamefont {Liberati}(2007)}]{liberati2007quantum}%
  \BibitemOpen
  \bibfield  {author} {\bibinfo {author} {\bibfnamefont {S.}~\bibnamefont {Liberati}},\ }\bibfield  {title} {\bibinfo {title} {Quantum gravity phenomenology via {Lorentz} violations},\ }\href@noop {} {\bibfield  {journal} {\bibinfo  {journal} {arXiv preprint arXiv:0706.0142}\ } (\bibinfo {year} {2007})}\BibitemShut {NoStop}%
\bibitem [{\citenamefont {Liberati}(2013)}]{liberati2013tests}%
  \BibitemOpen
  \bibfield  {author} {\bibinfo {author} {\bibfnamefont {S.}~\bibnamefont {Liberati}},\ }\bibfield  {title} {\bibinfo {title} {Tests of {Lorentz} invariance: a 2013 update},\ }\href@noop {} {\bibfield  {journal} {\bibinfo  {journal} {Class. Quant. Grav.}\ }\textbf {\bibinfo {volume} {30}},\ \bibinfo {pages} {133001} (\bibinfo {year} {2013})}\BibitemShut {NoStop}%
\bibitem [{\citenamefont {Barausse}\ \emph {et~al.}(2011)\citenamefont {Barausse}, \citenamefont {Jacobson},\ and\ \citenamefont {Sotiriou}}]{barausse2011black}%
  \BibitemOpen
  \bibfield  {author} {\bibinfo {author} {\bibfnamefont {E.}~\bibnamefont {Barausse}}, \bibinfo {author} {\bibfnamefont {T.}~\bibnamefont {Jacobson}},\ and\ \bibinfo {author} {\bibfnamefont {T.~P.}\ \bibnamefont {Sotiriou}},\ }\bibfield  {title} {\bibinfo {title} {Black holes in {Einstein-Aether} and {Ho{\v{r}}ava-Lifshitz} gravity},\ }\href@noop {} {\bibfield  {journal} {\bibinfo  {journal} {Phys. Rev. D}\ }\textbf {\bibinfo {volume} {83}},\ \bibinfo {pages} {124043} (\bibinfo {year} {2011})}\BibitemShut {NoStop}%
\bibitem [{\citenamefont {Barausse}\ and\ \citenamefont {Sotiriou}(2013)}]{barausse2013black}%
  \BibitemOpen
  \bibfield  {author} {\bibinfo {author} {\bibfnamefont {E.}~\bibnamefont {Barausse}}\ and\ \bibinfo {author} {\bibfnamefont {T.~P.}\ \bibnamefont {Sotiriou}},\ }\bibfield  {title} {\bibinfo {title} {Black holes in {Lorentz}-violating gravity theories},\ }\href@noop {} {\bibfield  {journal} {\bibinfo  {journal} {Class. Quant. Grav.}\ }\textbf {\bibinfo {volume} {30}},\ \bibinfo {pages} {244010} (\bibinfo {year} {2013})}\BibitemShut {NoStop}%
\bibitem [{\citenamefont {Barcelo}\ \emph {et~al.}(2011)\citenamefont {Barcelo}, \citenamefont {Liberati},\ and\ \citenamefont {Visser}}]{barcelo2011analogue}%
  \BibitemOpen
  \bibfield  {author} {\bibinfo {author} {\bibfnamefont {C.}~\bibnamefont {Barcelo}}, \bibinfo {author} {\bibfnamefont {S.}~\bibnamefont {Liberati}},\ and\ \bibinfo {author} {\bibfnamefont {M.}~\bibnamefont {Visser}},\ }\bibfield  {title} {\bibinfo {title} {Analogue gravity},\ }\href@noop {} {\bibfield  {journal} {\bibinfo  {journal} {Living Reviews in Relativity}\ }\textbf {\bibinfo {volume} {14}},\ \bibinfo {pages} {3} (\bibinfo {year} {2011})}\BibitemShut {NoStop}%
\bibitem [{\citenamefont {Unruh}(1981)}]{unruh1981experimental}%
  \BibitemOpen
  \bibfield  {author} {\bibinfo {author} {\bibfnamefont {W.~G.}\ \bibnamefont {Unruh}},\ }\bibfield  {title} {\bibinfo {title} {Experimental black-hole evaporation?},\ }\href@noop {} {\bibfield  {journal} {\bibinfo  {journal} {Physical Review Letters}\ }\textbf {\bibinfo {volume} {46}},\ \bibinfo {pages} {1351} (\bibinfo {year} {1981})}\BibitemShut {NoStop}%
\bibitem [{\citenamefont {Visser}(1993)}]{visser1993lorentzian}%
  \BibitemOpen
  \bibfield  {author} {\bibinfo {author} {\bibfnamefont {M.}~\bibnamefont {Visser}},\ }\bibfield  {title} {\bibinfo {title} {Acoustic propagation in fluids: An unexpected example of {Lorentzian} geometry},\ }\href@noop {} {\bibfield  {journal} {\bibinfo  {journal} {arXiv: gr-qc/9311028}\ } (\bibinfo {year} {1993})},\ \Eprint {https://arxiv.org/abs/gr-qc/9311028} {arXiv:gr-qc/9311028 [gr-qc]} \BibitemShut {NoStop}%
\bibitem [{\citenamefont {Unruh}(1995)}]{unruh1995sonic}%
  \BibitemOpen
  \bibfield  {author} {\bibinfo {author} {\bibfnamefont {W.~G.}\ \bibnamefont {Unruh}},\ }\bibfield  {title} {\bibinfo {title} {{Sonic analog of black holes and the effects of high frequencies on black hole evaporation}},\ }\href {https://doi.org/10.1103/PhysRevD.51.2827} {\bibfield  {journal} {\bibinfo  {journal} {Physical Review D}\ }\textbf {\bibinfo {volume} {51}},\ \bibinfo {pages} {2827} (\bibinfo {year} {1995})}\BibitemShut {NoStop}%
\bibitem [{\citenamefont {Corley}\ and\ \citenamefont {Jacobson}(1996)}]{corley1996spectrum}%
  \BibitemOpen
  \bibfield  {author} {\bibinfo {author} {\bibfnamefont {S.}~\bibnamefont {Corley}}\ and\ \bibinfo {author} {\bibfnamefont {T.}~\bibnamefont {Jacobson}},\ }\bibfield  {title} {\bibinfo {title} {{Hawking} spectrum and high frequency dispersion},\ }\href {https://doi.org/10.1103/PhysRevD.54.1568} {\bibfield  {journal} {\bibinfo  {journal} {Physical Review D}\ }\textbf {\bibinfo {volume} {54}},\ \bibinfo {pages} {1568} (\bibinfo {year} {1996})}\BibitemShut {NoStop}%
\bibitem [{\citenamefont {Rousseaux}\ \emph {et~al.}(2008)\citenamefont {Rousseaux}, \citenamefont {Mathis}, \citenamefont {Ma{\"\i}ssa}, \citenamefont {Philbin},\ and\ \citenamefont {Leonhardt}}]{rousseaux2008observation}%
  \BibitemOpen
  \bibfield  {author} {\bibinfo {author} {\bibfnamefont {G.}~\bibnamefont {Rousseaux}}, \bibinfo {author} {\bibfnamefont {C.}~\bibnamefont {Mathis}}, \bibinfo {author} {\bibfnamefont {P.}~\bibnamefont {Ma{\"\i}ssa}}, \bibinfo {author} {\bibfnamefont {T.~G.}\ \bibnamefont {Philbin}},\ and\ \bibinfo {author} {\bibfnamefont {U.}~\bibnamefont {Leonhardt}},\ }\bibfield  {title} {\bibinfo {title} {Observation of negative-frequency waves in a water tank: a classical analogue to the {Hawking} effect?},\ }\href@noop {} {\bibfield  {journal} {\bibinfo  {journal} {New Journal of Physics}\ }\textbf {\bibinfo {volume} {10}},\ \bibinfo {pages} {053015} (\bibinfo {year} {2008})}\BibitemShut {NoStop}%
\bibitem [{\citenamefont {Weinfurtner}\ \emph {et~al.}(2011)\citenamefont {Weinfurtner}, \citenamefont {Tedford}, \citenamefont {Penrice}, \citenamefont {Unruh},\ and\ \citenamefont {Lawrence}}]{weinfurtner2011measurement}%
  \BibitemOpen
  \bibfield  {author} {\bibinfo {author} {\bibfnamefont {S.}~\bibnamefont {Weinfurtner}}, \bibinfo {author} {\bibfnamefont {E.~W.}\ \bibnamefont {Tedford}}, \bibinfo {author} {\bibfnamefont {C.~J.}\ \bibnamefont {Penrice}, \bibfnamefont {M}}, \bibinfo {author} {\bibfnamefont {W.~G.}\ \bibnamefont {Unruh}},\ and\ \bibinfo {author} {\bibfnamefont {G.~A.}\ \bibnamefont {Lawrence}},\ }\bibfield  {title} {\bibinfo {title} {Measurement of stimulated {Hawking} emission in an analogue system},\ }\href@noop {} {\bibfield  {journal} {\bibinfo  {journal} {Phys. Rev. Lett}\ }\textbf {\bibinfo {volume} {106}},\ \bibinfo {pages} {021302} (\bibinfo {year} {2011})}\BibitemShut {NoStop}%
\bibitem [{\citenamefont {Euv{\'e}}\ \emph {et~al.}(2016)\citenamefont {Euv{\'e}}, \citenamefont {Michel}, \citenamefont {Parentani}, \citenamefont {Philbin},\ and\ \citenamefont {Rousseaux}}]{euve2016observation}%
  \BibitemOpen
  \bibfield  {author} {\bibinfo {author} {\bibfnamefont {L.~P.}\ \bibnamefont {Euv{\'e}}}, \bibinfo {author} {\bibfnamefont {F.}~\bibnamefont {Michel}}, \bibinfo {author} {\bibfnamefont {R.}~\bibnamefont {Parentani}}, \bibinfo {author} {\bibfnamefont {T.~G.}\ \bibnamefont {Philbin}},\ and\ \bibinfo {author} {\bibfnamefont {G.}~\bibnamefont {Rousseaux}},\ }\bibfield  {title} {\bibinfo {title} {Observation of noise correlated by the {Hawking} effect in a water tank},\ }\href@noop {} {\bibfield  {journal} {\bibinfo  {journal} {Physical Review Letters}\ }\textbf {\bibinfo {volume} {117}},\ \bibinfo {pages} {121301} (\bibinfo {year} {2016})}\BibitemShut {NoStop}%
\bibitem [{\citenamefont {Steinhauer}(2016)}]{steinhauer2016observation}%
  \BibitemOpen
  \bibfield  {author} {\bibinfo {author} {\bibfnamefont {J.}~\bibnamefont {Steinhauer}},\ }\bibfield  {title} {\bibinfo {title} {Observation of quantum hawking radiation and its entanglement in an analogue black hole},\ }\href@noop {} {\bibfield  {journal} {\bibinfo  {journal} {Nature Physics}\ }\textbf {\bibinfo {volume} {12}},\ \bibinfo {pages} {959} (\bibinfo {year} {2016})}\BibitemShut {NoStop}%
\bibitem [{\citenamefont {Mu{\~n}oz~de Nova}\ \emph {et~al.}(2019)\citenamefont {Mu{\~n}oz~de Nova}, \citenamefont {Golubkov}, \citenamefont {Kolobov},\ and\ \citenamefont {Steinhauer}}]{munoz2019observation}%
  \BibitemOpen
  \bibfield  {author} {\bibinfo {author} {\bibfnamefont {J.~R.}\ \bibnamefont {Mu{\~n}oz~de Nova}}, \bibinfo {author} {\bibfnamefont {K.}~\bibnamefont {Golubkov}}, \bibinfo {author} {\bibfnamefont {V.~I.}\ \bibnamefont {Kolobov}},\ and\ \bibinfo {author} {\bibfnamefont {J.}~\bibnamefont {Steinhauer}},\ }\bibfield  {title} {\bibinfo {title} {Observation of thermal {Hawking} radiation and its temperature in an analogue black hole},\ }\href@noop {} {\bibfield  {journal} {\bibinfo  {journal} {Nature}\ }\textbf {\bibinfo {volume} {569}},\ \bibinfo {pages} {688} (\bibinfo {year} {2019})}\BibitemShut {NoStop}%
\bibitem [{\citenamefont {Kolobov}\ \emph {et~al.}(2021)\citenamefont {Kolobov}, \citenamefont {Golubkov}, \citenamefont {Mu{\~n}oz~de Nova},\ and\ \citenamefont {Steinhauer}}]{kolobov2021observation}%
  \BibitemOpen
  \bibfield  {author} {\bibinfo {author} {\bibfnamefont {V.~I.}\ \bibnamefont {Kolobov}}, \bibinfo {author} {\bibfnamefont {K.}~\bibnamefont {Golubkov}}, \bibinfo {author} {\bibfnamefont {J.~R.}\ \bibnamefont {Mu{\~n}oz~de Nova}},\ and\ \bibinfo {author} {\bibfnamefont {J.}~\bibnamefont {Steinhauer}},\ }\bibfield  {title} {\bibinfo {title} {Observation of stationary spontaneous {Hawking} radiation and the time evolution of an analogue black hole},\ }\href@noop {} {\bibfield  {journal} {\bibinfo  {journal} {Nat. Phys.}\ }\textbf {\bibinfo {volume} {17}},\ \bibinfo {pages} {362} (\bibinfo {year} {2021})}\BibitemShut {NoStop}%
\bibitem [{\citenamefont {Torres}\ \emph {et~al.}(2017)\citenamefont {Torres}, \citenamefont {Patrick}, \citenamefont {Coutant}, \citenamefont {Richartz}, \citenamefont {Tedford},\ and\ \citenamefont {Weinfurtner}}]{torres2017rotational}%
  \BibitemOpen
  \bibfield  {author} {\bibinfo {author} {\bibfnamefont {T.}~\bibnamefont {Torres}}, \bibinfo {author} {\bibfnamefont {S.}~\bibnamefont {Patrick}}, \bibinfo {author} {\bibfnamefont {A.}~\bibnamefont {Coutant}}, \bibinfo {author} {\bibfnamefont {M.}~\bibnamefont {Richartz}}, \bibinfo {author} {\bibfnamefont {E.~W.}\ \bibnamefont {Tedford}},\ and\ \bibinfo {author} {\bibfnamefont {S.}~\bibnamefont {Weinfurtner}},\ }\bibfield  {title} {\bibinfo {title} {Rotational superradiant scattering in a vortex flow},\ }\href@noop {} {\bibfield  {journal} {\bibinfo  {journal} {Nature Physics}\ }\textbf {\bibinfo {volume} {13}},\ \bibinfo {pages} {833} (\bibinfo {year} {2017})}\BibitemShut {NoStop}%
\bibitem [{\citenamefont {Braidotti}\ \emph {et~al.}(2022)\citenamefont {Braidotti}, \citenamefont {Prizia}, \citenamefont {Maitland}, \citenamefont {Marino}, \citenamefont {Prain}, \citenamefont {Starshynov}, \citenamefont {Westerberg}, \citenamefont {Wright},\ and\ \citenamefont {Faccio}}]{braidotti2022measurement}%
  \BibitemOpen
  \bibfield  {author} {\bibinfo {author} {\bibfnamefont {M.~C.}\ \bibnamefont {Braidotti}}, \bibinfo {author} {\bibfnamefont {R.}~\bibnamefont {Prizia}}, \bibinfo {author} {\bibfnamefont {C.}~\bibnamefont {Maitland}}, \bibinfo {author} {\bibfnamefont {F.}~\bibnamefont {Marino}}, \bibinfo {author} {\bibfnamefont {A.}~\bibnamefont {Prain}}, \bibinfo {author} {\bibfnamefont {I.}~\bibnamefont {Starshynov}}, \bibinfo {author} {\bibfnamefont {N.}~\bibnamefont {Westerberg}}, \bibinfo {author} {\bibfnamefont {E.~M.}\ \bibnamefont {Wright}},\ and\ \bibinfo {author} {\bibfnamefont {D.}~\bibnamefont {Faccio}},\ }\bibfield  {title} {\bibinfo {title} {Measurement of penrose superradiance in a photon superfluid},\ }\href@noop {} {\bibfield  {journal} {\bibinfo  {journal} {Phys. Rev. Lett.}\ }\textbf {\bibinfo {volume} {128}},\ \bibinfo {pages} {013901} (\bibinfo {year} {2022})}\BibitemShut {NoStop}%
\bibitem [{\citenamefont {Patrick}\ and\ \citenamefont {Weinfurtner}(2020)}]{patrick2020superradiance}%
  \BibitemOpen
  \bibfield  {author} {\bibinfo {author} {\bibfnamefont {S.}~\bibnamefont {Patrick}}\ and\ \bibinfo {author} {\bibfnamefont {S.}~\bibnamefont {Weinfurtner}},\ }\bibfield  {title} {\bibinfo {title} {Superradiance in dispersive black hole analogues},\ }\href {https://doi.org/10.1103/PhysRevD.102.084041} {\bibfield  {journal} {\bibinfo  {journal} {Phys. Rev. D}\ }\textbf {\bibinfo {volume} {102}},\ \bibinfo {pages} {084041} (\bibinfo {year} {2020})}\BibitemShut {NoStop}%
\bibitem [{\citenamefont {Patrick}(2021)}]{patrick2021rotational}%
  \BibitemOpen
  \bibfield  {author} {\bibinfo {author} {\bibfnamefont {S.}~\bibnamefont {Patrick}},\ }\bibfield  {title} {\bibinfo {title} {Rotational superradiance with {Bogoliubov} dispersion},\ }\href@noop {} {\bibfield  {journal} {\bibinfo  {journal} {Class. Quant. Grav.}\ }\textbf {\bibinfo {volume} {38}},\ \bibinfo {pages} {095010} (\bibinfo {year} {2021})}\BibitemShut {NoStop}%
\bibitem [{\citenamefont {Patrick}\ and\ \citenamefont {Torres}(2024)}]{patrick2024primer}%
  \BibitemOpen
  \bibfield  {author} {\bibinfo {author} {\bibfnamefont {S.}~\bibnamefont {Patrick}}\ and\ \bibinfo {author} {\bibfnamefont {T.}~\bibnamefont {Torres}},\ }\bibfield  {title} {\bibinfo {title} {Primer on the analog black hole bomb with capillary-gravity waves},\ }\href {https://doi.org/10.1103/PhysRevD.110.124068} {\bibfield  {journal} {\bibinfo  {journal} {Phys. Rev. D}\ }\textbf {\bibinfo {volume} {110}},\ \bibinfo {pages} {124068} (\bibinfo {year} {2024})}\BibitemShut {NoStop}%
\bibitem [{\citenamefont {Torres}\ \emph {et~al.}(2020)\citenamefont {Torres}, \citenamefont {Patrick}, \citenamefont {Richartz},\ and\ \citenamefont {Weinfurtner}}]{torres2020quasinormal}%
  \BibitemOpen
  \bibfield  {author} {\bibinfo {author} {\bibfnamefont {T.}~\bibnamefont {Torres}}, \bibinfo {author} {\bibfnamefont {S.}~\bibnamefont {Patrick}}, \bibinfo {author} {\bibfnamefont {M.}~\bibnamefont {Richartz}},\ and\ \bibinfo {author} {\bibfnamefont {S.}~\bibnamefont {Weinfurtner}},\ }\bibfield  {title} {\bibinfo {title} {Quasinormal mode oscillations in an analogue black hole experiment},\ }\href {https://doi.org/10.1103/PhysRevLett.125.011301} {\bibfield  {journal} {\bibinfo  {journal} {Phys. Rev. Lett.}\ }\textbf {\bibinfo {volume} {125}},\ \bibinfo {pages} {011301} (\bibinfo {year} {2020})}\BibitemShut {NoStop}%
\bibitem [{\citenamefont {Smaniotto}\ \emph {et~al.}(2025)\citenamefont {Smaniotto}, \citenamefont {Solidoro}, \citenamefont {{\v{S}}van{\v{c}}ara}, \citenamefont {Patrick}, \citenamefont {Richartz}, \citenamefont {Barenghi}, \citenamefont {Gregory},\ and\ \citenamefont {Weinfurtner}}]{smaniotto2025black}%
  \BibitemOpen
  \bibfield  {author} {\bibinfo {author} {\bibfnamefont {P.}~\bibnamefont {Smaniotto}}, \bibinfo {author} {\bibfnamefont {L.}~\bibnamefont {Solidoro}}, \bibinfo {author} {\bibfnamefont {P.}~\bibnamefont {{\v{S}}van{\v{c}}ara}}, \bibinfo {author} {\bibfnamefont {S.}~\bibnamefont {Patrick}}, \bibinfo {author} {\bibfnamefont {M.}~\bibnamefont {Richartz}}, \bibinfo {author} {\bibfnamefont {C.~F.}\ \bibnamefont {Barenghi}}, \bibinfo {author} {\bibfnamefont {R.}~\bibnamefont {Gregory}},\ and\ \bibinfo {author} {\bibfnamefont {S.}~\bibnamefont {Weinfurtner}},\ }\bibfield  {title} {\bibinfo {title} {Black-hole spectroscopy from a giant quantum vortex},\ }\href@noop {} {\bibfield  {journal} {\bibinfo  {journal} {arXiv preprint arXiv:2502.11209}\ } (\bibinfo {year} {2025})}\BibitemShut {NoStop}%
\bibitem [{\citenamefont {{\v{S}}van{\v{c}}ara}\ \emph {et~al.}(2024)\citenamefont {{\v{S}}van{\v{c}}ara}, \citenamefont {Smaniotto}, \citenamefont {Solidoro}, \citenamefont {MacDonald}, \citenamefont {Patrick}, \citenamefont {Gregory}, \citenamefont {Barenghi},\ and\ \citenamefont {Weinfurtner}}]{svanvcara2024rotating}%
  \BibitemOpen
  \bibfield  {author} {\bibinfo {author} {\bibfnamefont {P.}~\bibnamefont {{\v{S}}van{\v{c}}ara}}, \bibinfo {author} {\bibfnamefont {P.}~\bibnamefont {Smaniotto}}, \bibinfo {author} {\bibfnamefont {L.}~\bibnamefont {Solidoro}}, \bibinfo {author} {\bibfnamefont {J.~F.}\ \bibnamefont {MacDonald}}, \bibinfo {author} {\bibfnamefont {S.}~\bibnamefont {Patrick}}, \bibinfo {author} {\bibfnamefont {R.}~\bibnamefont {Gregory}}, \bibinfo {author} {\bibfnamefont {C.~F.}\ \bibnamefont {Barenghi}},\ and\ \bibinfo {author} {\bibfnamefont {S.}~\bibnamefont {Weinfurtner}},\ }\bibfield  {title} {\bibinfo {title} {Rotating curved spacetime signatures from a giant quantum vortex},\ }\href@noop {} {\bibfield  {journal} {\bibinfo  {journal} {Nature}\ }\textbf {\bibinfo {volume} {628}},\ \bibinfo {pages} {66} (\bibinfo {year} {2024})}\BibitemShut {NoStop}%
\bibitem [{\citenamefont {Delhom}\ \emph {et~al.}(2024)\citenamefont {Delhom}, \citenamefont {Guerrero}, \citenamefont {Calizaya~Cabrera}, \citenamefont {Falque}, \citenamefont {Bramati}, \citenamefont {Brady}, \citenamefont {Jacquet},\ and\ \citenamefont {Agullo}}]{delhom2024entanglement}%
  \BibitemOpen
  \bibfield  {author} {\bibinfo {author} {\bibfnamefont {A.}~\bibnamefont {Delhom}}, \bibinfo {author} {\bibfnamefont {K.}~\bibnamefont {Guerrero}}, \bibinfo {author} {\bibfnamefont {P.}~\bibnamefont {Calizaya~Cabrera}}, \bibinfo {author} {\bibfnamefont {K.}~\bibnamefont {Falque}}, \bibinfo {author} {\bibfnamefont {A.}~\bibnamefont {Bramati}}, \bibinfo {author} {\bibfnamefont {A.~J.}\ \bibnamefont {Brady}}, \bibinfo {author} {\bibfnamefont {M.~J.}\ \bibnamefont {Jacquet}},\ and\ \bibinfo {author} {\bibfnamefont {I.}~\bibnamefont {Agullo}},\ }\bibfield  {title} {\bibinfo {title} {Entanglement from superradiance and rotating quantum fluids of light},\ }\href@noop {} {\bibfield  {journal} {\bibinfo  {journal} {Phys. Rev. D}\ }\textbf {\bibinfo {volume} {109}},\ \bibinfo {pages} {105024} (\bibinfo {year} {2024})}\BibitemShut {NoStop}%
\bibitem [{\citenamefont {Torres}(2020)}]{torres2020estimate}%
  \BibitemOpen
  \bibfield  {author} {\bibinfo {author} {\bibfnamefont {T.}~\bibnamefont {Torres}},\ }\bibfield  {title} {\bibinfo {title} {Estimate of the superradiance spectrum in dispersive media},\ }\href@noop {} {\bibfield  {journal} {\bibinfo  {journal} {Philosophical Transactions of the Royal Society A}\ }\textbf {\bibinfo {volume} {378}},\ \bibinfo {pages} {20190236} (\bibinfo {year} {2020})}\BibitemShut {NoStop}%
\bibitem [{\citenamefont {Iyer}\ and\ \citenamefont {Will}(1987)}]{iyer1987black1}%
  \BibitemOpen
  \bibfield  {author} {\bibinfo {author} {\bibfnamefont {S.}~\bibnamefont {Iyer}}\ and\ \bibinfo {author} {\bibfnamefont {C.~M.}\ \bibnamefont {Will}},\ }\bibfield  {title} {\bibinfo {title} {Black-hole normal modes: A {WKB} approach. {I. Foundations} and application of a higher-order {WKB} analysis of potential-barrier scattering},\ }\href@noop {} {\bibfield  {journal} {\bibinfo  {journal} {Physical Review D}\ }\textbf {\bibinfo {volume} {35}},\ \bibinfo {pages} {3621} (\bibinfo {year} {1987})}\BibitemShut {NoStop}%
\bibitem [{\citenamefont {Torres}\ \emph {et~al.}(2018)\citenamefont {Torres}, \citenamefont {Coutant}, \citenamefont {Dolan},\ and\ \citenamefont {Weinfurtner}}]{torres2018waves}%
  \BibitemOpen
  \bibfield  {author} {\bibinfo {author} {\bibfnamefont {T.}~\bibnamefont {Torres}}, \bibinfo {author} {\bibfnamefont {A.}~\bibnamefont {Coutant}}, \bibinfo {author} {\bibfnamefont {S.}~\bibnamefont {Dolan}},\ and\ \bibinfo {author} {\bibfnamefont {S.}~\bibnamefont {Weinfurtner}},\ }\bibfield  {title} {\bibinfo {title} {Waves on a vortex: rays, rings and resonances},\ }\href@noop {} {\bibfield  {journal} {\bibinfo  {journal} {Journal of Fluid Mechanics}\ }\textbf {\bibinfo {volume} {857}},\ \bibinfo {pages} {291} (\bibinfo {year} {2018})}\BibitemShut {NoStop}%
\bibitem [{\citenamefont {Cardoso}\ \emph {et~al.}(2009)\citenamefont {Cardoso}, \citenamefont {Miranda}, \citenamefont {Berti}, \citenamefont {Witek},\ and\ \citenamefont {Zanchin}}]{cardoso2009geodesic}%
  \BibitemOpen
  \bibfield  {author} {\bibinfo {author} {\bibfnamefont {V.}~\bibnamefont {Cardoso}}, \bibinfo {author} {\bibfnamefont {A.~S.}\ \bibnamefont {Miranda}}, \bibinfo {author} {\bibfnamefont {E.}~\bibnamefont {Berti}}, \bibinfo {author} {\bibfnamefont {H.}~\bibnamefont {Witek}},\ and\ \bibinfo {author} {\bibfnamefont {V.~T.}\ \bibnamefont {Zanchin}},\ }\bibfield  {title} {\bibinfo {title} {Geodesic stability, {Lyapunov} exponents, and quasinormal modes},\ }\href@noop {} {\bibfield  {journal} {\bibinfo  {journal} {Physical Review D}\ }\textbf {\bibinfo {volume} {79}},\ \bibinfo {pages} {064016} (\bibinfo {year} {2009})}\BibitemShut {NoStop}%
\bibitem [{\citenamefont {Konoplya}\ \emph {et~al.}(2019)\citenamefont {Konoplya}, \citenamefont {Zhidenko},\ and\ \citenamefont {Zinhailo}}]{konoplya2019higher}%
  \BibitemOpen
  \bibfield  {author} {\bibinfo {author} {\bibfnamefont {R.~A.}\ \bibnamefont {Konoplya}}, \bibinfo {author} {\bibfnamefont {A.}~\bibnamefont {Zhidenko}},\ and\ \bibinfo {author} {\bibfnamefont {A.~F.}\ \bibnamefont {Zinhailo}},\ }\bibfield  {title} {\bibinfo {title} {Higher order {WKB} formula for quasinormal modes and grey-body factors: recipes for quick and accurate calculations},\ }\href@noop {} {\bibfield  {journal} {\bibinfo  {journal} {Class. Quant. Grav.}\ }\textbf {\bibinfo {volume} {36}},\ \bibinfo {pages} {155002} (\bibinfo {year} {2019})}\BibitemShut {NoStop}%
\bibitem [{\citenamefont {Matyjasek}\ \emph {et~al.}(2024)\citenamefont {Matyjasek}, \citenamefont {Benda},\ and\ \citenamefont {Stafi{\'n}ska}}]{matyjasek2024accurate}%
  \BibitemOpen
  \bibfield  {author} {\bibinfo {author} {\bibfnamefont {J.}~\bibnamefont {Matyjasek}}, \bibinfo {author} {\bibfnamefont {K.}~\bibnamefont {Benda}},\ and\ \bibinfo {author} {\bibfnamefont {M.}~\bibnamefont {Stafi{\'n}ska}},\ }\bibfield  {title} {\bibinfo {title} {Accurate quasinormal modes of the analog black holes},\ }\href@noop {} {\bibfield  {journal} {\bibinfo  {journal} {Phys. Rev. D}\ }\textbf {\bibinfo {volume} {110}},\ \bibinfo {pages} {064083} (\bibinfo {year} {2024})}\BibitemShut {NoStop}%
\bibitem [{\citenamefont {Cardoso}\ \emph {et~al.}(2004)\citenamefont {Cardoso}, \citenamefont {Lemos},\ and\ \citenamefont {Yoshida}}]{cardoso2004qnm}%
  \BibitemOpen
  \bibfield  {author} {\bibinfo {author} {\bibfnamefont {V.}~\bibnamefont {Cardoso}}, \bibinfo {author} {\bibfnamefont {J.~P.~S.}\ \bibnamefont {Lemos}},\ and\ \bibinfo {author} {\bibfnamefont {S.}~\bibnamefont {Yoshida}},\ }\bibfield  {title} {\bibinfo {title} {Quasinormal modes and stability of the rotating acoustic black hole: Numerical analysis},\ }\href@noop {} {\bibfield  {journal} {\bibinfo  {journal} {Physical Review D}\ }\textbf {\bibinfo {volume} {70}},\ \bibinfo {pages} {124032} (\bibinfo {year} {2004})}\BibitemShut {NoStop}%
\bibitem [{\citenamefont {Berti}\ \emph {et~al.}(2004)\citenamefont {Berti}, \citenamefont {Cardoso},\ and\ \citenamefont {Lemos}}]{berti2004qnm}%
  \BibitemOpen
  \bibfield  {author} {\bibinfo {author} {\bibfnamefont {E.}~\bibnamefont {Berti}}, \bibinfo {author} {\bibfnamefont {V.}~\bibnamefont {Cardoso}},\ and\ \bibinfo {author} {\bibfnamefont {J.~P.~S.}\ \bibnamefont {Lemos}},\ }\bibfield  {title} {\bibinfo {title} {Quasinormal modes and classical wave propagation in analogue black holes},\ }\href@noop {} {\bibfield  {journal} {\bibinfo  {journal} {Physical Review D}\ }\textbf {\bibinfo {volume} {70}},\ \bibinfo {pages} {124006} (\bibinfo {year} {2004})}\BibitemShut {NoStop}%
\bibitem [{\citenamefont {Solidoro}\ \emph {et~al.}(2024)\citenamefont {Solidoro}, \citenamefont {Patrick}, \citenamefont {Gregory},\ and\ \citenamefont {Weinfurtner}}]{solidoro2024quasinormal}%
  \BibitemOpen
  \bibfield  {author} {\bibinfo {author} {\bibfnamefont {L.}~\bibnamefont {Solidoro}}, \bibinfo {author} {\bibfnamefont {S.}~\bibnamefont {Patrick}}, \bibinfo {author} {\bibfnamefont {R.}~\bibnamefont {Gregory}},\ and\ \bibinfo {author} {\bibfnamefont {S.}~\bibnamefont {Weinfurtner}},\ }\bibfield  {title} {\bibinfo {title} {Quasinormal modes in semi-open systems},\ }\href@noop {} {\bibfield  {journal} {\bibinfo  {journal} {arXiv preprint arXiv:2406.11013}\ } (\bibinfo {year} {2024})}\BibitemShut {NoStop}%
\bibitem [{\citenamefont {Visser}(1998)}]{visser1998acoustic}%
  \BibitemOpen
  \bibfield  {author} {\bibinfo {author} {\bibfnamefont {M.}~\bibnamefont {Visser}},\ }\bibfield  {title} {\bibinfo {title} {Acoustic black holes: horizons, ergospheres and {Hawking} radiation},\ }\href@noop {} {\bibfield  {journal} {\bibinfo  {journal} {Classical and Quantum Gravity}\ }\textbf {\bibinfo {volume} {15}},\ \bibinfo {pages} {1767} (\bibinfo {year} {1998})}\BibitemShut {NoStop}%
\bibitem [{\citenamefont {Matyjasek}(2021)}]{matyjasek2021accurate}%
  \BibitemOpen
  \bibfield  {author} {\bibinfo {author} {\bibfnamefont {J.}~\bibnamefont {Matyjasek}},\ }\bibfield  {title} {\bibinfo {title} {Accurate quasinormal modes of the five-dimensional {Schwarzschild-Tangherlini} black holes},\ }\href@noop {} {\bibfield  {journal} {\bibinfo  {journal} {Phys. Rev. D}\ }\textbf {\bibinfo {volume} {104}},\ \bibinfo {pages} {084066} (\bibinfo {year} {2021})}\BibitemShut {NoStop}%
\bibitem [{Note1()}]{Note1}%
  \BibitemOpen
  \bibinfo {note} {Note that in a quantum description, \protect \eqref {disp1} is related to \protect \eqref {E-M} in the rest frame of the fluid by a factor $\hbar ^2$, in which case it gives the $E-P$ relation for quantised collective excitations of the fluid}\BibitemShut {NoStop}%
\bibitem [{\citenamefont {Purser}\ and\ \citenamefont {Synge}(1962)}]{purser1962water}%
  \BibitemOpen
  \bibfield  {author} {\bibinfo {author} {\bibfnamefont {W.~F.~C.}\ \bibnamefont {Purser}}\ and\ \bibinfo {author} {\bibfnamefont {J.~L.}\ \bibnamefont {Synge}},\ }\bibfield  {title} {\bibinfo {title} {Water waves and {Hamilton's} method},\ }\href@noop {} {\bibfield  {journal} {\bibinfo  {journal} {Nature}\ }\textbf {\bibinfo {volume} {194}},\ \bibinfo {pages} {268} (\bibinfo {year} {1962})}\BibitemShut {NoStop}%
\bibitem [{Note2()}]{Note2}%
  \BibitemOpen
  \bibinfo {note} {Since energy is the product of the frequency and wave norm, negative energy solutions are allowed whenever $\omega >0$ are on the lower branch of the dispersion relation. This is the key mechanism underpinning superradiant scattering in rotating black holes and their analogues \cite {patrick2022quantum}. In our discussion of quasinormal modes, however, this mechanism will not play a role.}\BibitemShut {Stop}%
\bibitem [{Note3()}]{Note3}%
  \BibitemOpen
  \bibinfo {note} {This would not be the case, for example, if the out-going mode were reflected back into the vortex by a combination of a dispersion and velocity profile.}\BibitemShut {Stop}%
\bibitem [{\citenamefont {Leaver}(1985)}]{leaver1985analytic}%
  \BibitemOpen
  \bibfield  {author} {\bibinfo {author} {\bibfnamefont {E.~W.}\ \bibnamefont {Leaver}},\ }\bibfield  {title} {\bibinfo {title} {An analytic representation for the quasi-normal modes of {Kerr} black holes},\ }\href {https://doi.org/10.1098/rspa.1985.0119} {\bibfield  {journal} {\bibinfo  {journal} {Proceedings of the Royal Society of London}\ }\textbf {\bibinfo {volume} {A 402}},\ \bibinfo {pages} {285} (\bibinfo {year} {1985})}\BibitemShut {NoStop}%
\bibitem [{\citenamefont {Burgess}\ \emph {et~al.}(2024)\citenamefont {Burgess}, \citenamefont {Patrick}, \citenamefont {Torres}, \citenamefont {Gregory},\ and\ \citenamefont {K{\"o}nig}}]{burgess2024quasinormal}%
  \BibitemOpen
  \bibfield  {author} {\bibinfo {author} {\bibfnamefont {C.}~\bibnamefont {Burgess}}, \bibinfo {author} {\bibfnamefont {S.}~\bibnamefont {Patrick}}, \bibinfo {author} {\bibfnamefont {T.}~\bibnamefont {Torres}}, \bibinfo {author} {\bibfnamefont {R.}~\bibnamefont {Gregory}},\ and\ \bibinfo {author} {\bibfnamefont {F.}~\bibnamefont {K{\"o}nig}},\ }\bibfield  {title} {\bibinfo {title} {Quasinormal modes of optical solitons},\ }\href@noop {} {\bibfield  {journal} {\bibinfo  {journal} {Phys. Rev. Lett.}\ }\textbf {\bibinfo {volume} {132}},\ \bibinfo {pages} {053802} (\bibinfo {year} {2024})}\BibitemShut {NoStop}%
\bibitem [{\citenamefont {Shin}\ \emph {et~al.}(2004)\citenamefont {Shin}, \citenamefont {Saba}, \citenamefont {Vengalattore}, \citenamefont {Pasquini}, \citenamefont {Sanner}, \citenamefont {Leanhardt}, \citenamefont {Prentiss}, \citenamefont {Pritchard},\ and\ \citenamefont {Ketterle}}]{shin2004dynamical}%
  \BibitemOpen
  \bibfield  {author} {\bibinfo {author} {\bibfnamefont {Y.}~\bibnamefont {Shin}}, \bibinfo {author} {\bibfnamefont {M.}~\bibnamefont {Saba}}, \bibinfo {author} {\bibfnamefont {M.}~\bibnamefont {Vengalattore}}, \bibinfo {author} {\bibfnamefont {T.~A.}\ \bibnamefont {Pasquini}}, \bibinfo {author} {\bibfnamefont {C.}~\bibnamefont {Sanner}}, \bibinfo {author} {\bibfnamefont {A.~E.}\ \bibnamefont {Leanhardt}}, \bibinfo {author} {\bibfnamefont {M.}~\bibnamefont {Prentiss}}, \bibinfo {author} {\bibfnamefont {D.~E.}\ \bibnamefont {Pritchard}},\ and\ \bibinfo {author} {\bibfnamefont {W.}~\bibnamefont {Ketterle}},\ }\bibfield  {title} {\bibinfo {title} {Dynamical instability of a doubly quantized vortex in a {Bose-Einstein} condensate},\ }\href@noop {} {\bibfield  {journal} {\bibinfo  {journal} {Phys. Rev. Lett.}\ }\textbf {\bibinfo {volume} {93}},\ \bibinfo {pages} {160406} (\bibinfo {year} {2004})}\BibitemShut {NoStop}%
\bibitem [{\citenamefont {Giacomelli}\ and\ \citenamefont {Carusotto}(2020)}]{giacomelli2020ergoregion}%
  \BibitemOpen
  \bibfield  {author} {\bibinfo {author} {\bibfnamefont {L.}~\bibnamefont {Giacomelli}}\ and\ \bibinfo {author} {\bibfnamefont {I.}~\bibnamefont {Carusotto}},\ }\bibfield  {title} {\bibinfo {title} {Ergoregion instabilities in rotating two-dimensional {Bose-Einstein} condensates: {Perspectives} on the stability of quantized vortices},\ }\href@noop {} {\bibfield  {journal} {\bibinfo  {journal} {Phy. Rev. Research}\ }\textbf {\bibinfo {volume} {2}},\ \bibinfo {pages} {033139} (\bibinfo {year} {2020})}\BibitemShut {NoStop}%
\bibitem [{\citenamefont {Patrick}\ \emph {et~al.}(2022)\citenamefont {Patrick}, \citenamefont {Geelmuyden}, \citenamefont {Erne}, \citenamefont {Barenghi},\ and\ \citenamefont {Weinfurtner}}]{patrick2022quantum}%
  \BibitemOpen
  \bibfield  {author} {\bibinfo {author} {\bibfnamefont {S.}~\bibnamefont {Patrick}}, \bibinfo {author} {\bibfnamefont {A.}~\bibnamefont {Geelmuyden}}, \bibinfo {author} {\bibfnamefont {S.}~\bibnamefont {Erne}}, \bibinfo {author} {\bibfnamefont {C.~F.}\ \bibnamefont {Barenghi}},\ and\ \bibinfo {author} {\bibfnamefont {S.}~\bibnamefont {Weinfurtner}},\ }\bibfield  {title} {\bibinfo {title} {Quantum vortex instability and black hole superradiance},\ }\href@noop {} {\bibfield  {journal} {\bibinfo  {journal} {Phys. Rev. Research}\ }\textbf {\bibinfo {volume} {4}},\ \bibinfo {pages} {033117} (\bibinfo {year} {2022})}\BibitemShut {NoStop}%
\bibitem [{\citenamefont {Patrick}(2024)}]{patrick2024quantum}%
  \BibitemOpen
  \bibfield  {author} {\bibinfo {author} {\bibfnamefont {S.}~\bibnamefont {Patrick}},\ }\bibfield  {title} {\bibinfo {title} {Quantum vortex stability in draining fluid flows},\ }\href@noop {} {\bibfield  {journal} {\bibinfo  {journal} {Phys. Rev. A}\ }\textbf {\bibinfo {volume} {110}},\ \bibinfo {pages} {013327} (\bibinfo {year} {2024})}\BibitemShut {NoStop}%
\bibitem [{\citenamefont {Hod}(1998)}]{hod1998bohr}%
  \BibitemOpen
  \bibfield  {author} {\bibinfo {author} {\bibfnamefont {S.}~\bibnamefont {Hod}},\ }\bibfield  {title} {\bibinfo {title} {Bohr's correspondence principle and the area spectrum of quantum black holes},\ }\href@noop {} {\bibfield  {journal} {\bibinfo  {journal} {Phys. Rev. Lett.}\ }\textbf {\bibinfo {volume} {81}},\ \bibinfo {pages} {4293} (\bibinfo {year} {1998})}\BibitemShut {NoStop}%
\bibitem [{\citenamefont {Maggiore}(2008)}]{maggiore2008physical}%
  \BibitemOpen
  \bibfield  {author} {\bibinfo {author} {\bibfnamefont {M.}~\bibnamefont {Maggiore}},\ }\bibfield  {title} {\bibinfo {title} {Physical interpretation of the spectrum of black hole quasinormal modes},\ }\href@noop {} {\bibfield  {journal} {\bibinfo  {journal} {Phys. Rev. Lett.}\ }\textbf {\bibinfo {volume} {100}},\ \bibinfo {pages} {141301} (\bibinfo {year} {2008})}\BibitemShut {NoStop}%
\bibitem [{\citenamefont {Carr}\ \emph {et~al.}(2021)\citenamefont {Carr}, \citenamefont {Kohri}, \citenamefont {Sendouda},\ and\ \citenamefont {Yokoyama}}]{carr2021constraints}%
  \BibitemOpen
  \bibfield  {author} {\bibinfo {author} {\bibfnamefont {B.}~\bibnamefont {Carr}}, \bibinfo {author} {\bibfnamefont {K.}~\bibnamefont {Kohri}}, \bibinfo {author} {\bibfnamefont {Y.}~\bibnamefont {Sendouda}},\ and\ \bibinfo {author} {\bibfnamefont {J.}~\bibnamefont {Yokoyama}},\ }\bibfield  {title} {\bibinfo {title} {Constraints on primordial black holes},\ }\href@noop {} {\bibfield  {journal} {\bibinfo  {journal} {Rep. Prog. Phys.}\ }\textbf {\bibinfo {volume} {84}},\ \bibinfo {pages} {116902} (\bibinfo {year} {2021})}\BibitemShut {NoStop}%
\bibitem [{Note4()}]{Note4}%
  \BibitemOpen
  \bibinfo {note} {Note that for $\omega \in \protect \mathbb {C}$, the prefactors in \protect \eqref {bXK_def} make $X$ complex. However, if the imaginary part is small then the phase of the arguments only receive a small correction. In that case, the asymptotic expansions derived for $\omega \in \protect \mathbb {R}$ (which apply in a particular region of the complex plane) will still be valid, since a perturbative $\protect \mathrm {Im}[\omega ]$ will not displace the function out of this region.}\BibitemShut {Stop}%
\bibitem [{\citenamefont {Churilov}\ and\ \citenamefont {Stepanyants}(2019)}]{churilov2018scattering}%
  \BibitemOpen
  \bibfield  {author} {\bibinfo {author} {\bibfnamefont {S.}~\bibnamefont {Churilov}}\ and\ \bibinfo {author} {\bibfnamefont {Y.}~\bibnamefont {Stepanyants}},\ }\bibfield  {title} {\bibinfo {title} {Scattering of surface shallow water waves on a draining bathtub vortex},\ }\href@noop {} {\bibfield  {journal} {\bibinfo  {journal} {Physical Review Fluids}\ }\textbf {\bibinfo {volume} {4}},\ \bibinfo {pages} {034704} (\bibinfo {year} {2019})}\BibitemShut {NoStop}%
\bibitem [{\citenamefont {Patrick}(2020)}]{patrick2020analogy}%
  \BibitemOpen
  \bibfield  {author} {\bibinfo {author} {\bibfnamefont {S.}~\bibnamefont {Patrick}},\ }\bibfield  {title} {\bibinfo {title} {On the analogy between black holes and bathtub vortices},\ }\href@noop {} {\bibfield  {journal} {\bibinfo  {journal} {arXiv preprint arXiv:2009.02133}\ } (\bibinfo {year} {2020})}\BibitemShut {NoStop}%
\end{thebibliography}%
\end{document}